\documentstyle[12pt]{article}
\begin{document}

\def\ds{\displaystyle}
\def\beq{\begin{equation}}
\def\eeq{\end{equation}}
\def\bea{\begin{eqnarray}}
\def\eea{\end{eqnarray}}
\def\beeq{\begin{eqnarray}}
\def\eeeq{\end{eqnarray}}
\def\ve{\vert}
\def\vel{\left|}
\def\ver{\right|}
\def\nnb{\nonumber}
\def\ga{\left(}
\def\dr{\right)}
\def\aga{\left\{}
\def\adr{\right\}}
\def\lla{\left<}
\def\rra{\right>}
\def\rar{\rightarrow}
\def\nnb{\nonumber}
\def\la{\langle}
\def\ra{\rangle}
\def\ba{\begin{array}}
\def\ea{\end{array}}
\def\tr{\mbox{Tr}}
\def\ssp{{\Sigma^{*+}}}
\def\sso{{\Sigma^{*0}}}
\def\ssm{{\Sigma^{*-}}}
\def\xis0{{\Xi^{*0}}}
\def\xism{{\Xi^{*-}}}
\def\qs{\la \bar s s \ra}
\def\qu{\la \bar u u \ra}
\def\qd{\la \bar d d \ra}
\def\qq{\la \bar q q \ra}
\def\gGgG{\la g^2 G^2 \ra}
\def\q{\gamma_5 \not\!q}
\def\x{\gamma_5 \not\!x}
\def\g5{\gamma_5}
\def\sb{S_Q^{cf}}
\def\sd{S_d^{be}}
\def\su{S_u^{ad}}
\def\ss{S_s^{??}}
\def\sbp{{S}_Q^{'cf}}
\def\sdp{{S}_d^{'be}}
\def\sup{{S}_u^{'ad}}
\def\ssp{{S}_s^{'??}}
\def\sig{\sigma_{\mu \nu} \gamma_5 p^\mu q^\nu}
\def\fo{f_0(\frac{s_0}{M^2})}
\def\ffi{f_1(\frac{s_0}{M^2})}
\def\fii{f_2(\frac{s_0}{M^2})}
\def\O{{\cal O}}
\def\sl{{\Sigma^0 \Lambda}}
\def\es{\!\!\! &=& \!\!\!}
\def\ap{\!\!\! &\approx& \!\!\!}
\def\ar{&+& \!\!\!}
\def\ek{&-& \!\!\!}
\def\kek{\!\!\!&-& \!\!\!}
\def\cp{&\times& \!\!\!}
\def\se{\!\!\! &\simeq& \!\!\!}
\def\eqv{&\equiv& \!\!\!}
\def\kpm{&\pm& \!\!\!}
\def\kmp{&\mp& \!\!\!}


\def\simlt{\stackrel{<}{{}_\sim}}
\def\simgt{\stackrel{>}{{}_\sim}}


\title{
         {\Large
                 {\bf
Forward--backward asymmetries in 
$\Lambda_b \rar \Lambda \ell^+ \ell^-$ decay beyond the standard model 
                 }
         }
      }

\author{\vspace{1cm}\\
{\small T. M. Aliev \thanks
{e-mail: taliev@metu.edu.tr}\,\,,
V. Bashiry
\,\,,
M. Savc{\i} \thanks
{e-mail: savci@metu.edu.tr}} \\
{\small Physics Department, Middle East Technical University,
06531 Ankara, Turkey} }

\date{}

\begin{titlepage}
\maketitle
\thispagestyle{empty}

\begin{abstract}
We study the doubly--polarized lepton pair forward--backward asymmetries in
$\Lambda_b \rar \Lambda \ell^+ \ell^-$ decay using a general, model 
independent form of the effective Hamiltonian. We present the general 
expression for nine double--polarization forward--backward asymmetries. 
It is observed that, the study of the forward--backward asymmetries of the 
doubly--polarized lepton pair is a very useful tool for establishing 
new physics beyond the standard model. Moreover, the correlation between
forward--backward asymmetry and branching ratio is investigated. It is shown
that there exist certain certain regions of the new Wilson coefficients
where new physics can be established by measuring the polarized
forward--backward asymmetry only.  
\end{abstract}

~~~PACS numbers: 12.60.--i, 13.30.--a. 13.88.+e
\end{titlepage}

\section{Introduction}

With the B--factories at work, search of rare decays induced by the  
flavor--changing neutral current (FCNC) $b \rar s(d) \ell^+ \ell^-$ 
is now entering a new interesting era. These transitions provide an 
important consistency check of the standard model (SM) at loop 
level, since FCNC transitions are forbidden in the SM at tree level.    
These decays induced by the FCNC are very sensitive to the
new physics beyond the SM. New physics appear in rare
decays through the Wilson coefficients which can take values different from
their SM counterpart or through the new operator structures in an effective
Hamiltonian.

First measurements of the $B \rar X_s \gamma$ decay were reported by CLEO
Collaboration \cite{R6501} and more precise measurements are
currently being carried out in the experiments at B factories (see for
example \cite{R6502}). Exclusive decay involving the $b \rar s \gamma$ 
transition has been measured in \cite{R6503,R6504}. After these 
measurements of the radiative decay induced by the $b \rar s \gamma$ 
transition, main interest has been focused on the rare decays induced 
by the $b \rar s \ell^+ \ell^-$ transitions, which have relatively large 
branching ratio in the SM. These decays have been extensively studied 
in the SM and its various extensions \cite{R6505}--\cite{R6513}.

The exclusive $B \rar K^\ast (K) \ell^+ \ell^-$ decays, which are described
by $b \rar s \ell^+ \ell^-$ transition at quark level, have been 
studied comprehensively in literature (see \cite{R6511}--\cite{R6517} and 
references therein). Recently, BELLE and BaBar Collaborations announced the
following results for the branching ratios of both decays:
\bea
{\cal B}(B \rar K^\ast \ell^+ \ell^-) = \left\{ \begin{array}{lc}
\left( 11.5^{+2.6}_{-2.4} \pm 0.8 \pm 0.2\right) \times
10^{-7}& \cite{R6518}~,\\ \\
\left( 0.88^{+0.33}_{-0.29} \right) \times
10^{-6}& \cite{R6519}~,\end{array} \right. \nnb
\eea
\bea
{\cal B}(B \rar K \ell^+ \ell^-) = \left\{ \begin{array}{lc}
\left( 4.8^{+1.0}_{-0.9} \pm 0.3 \pm 0.1\right) \times
10^{-7}& \cite{R6518}~,\\ \\
\left( 0.65^{+0.14}_{-0.13} \pm 0.04 \right) \times
10^{-6}& \cite{R6519}~.\end{array} \right. \nnb
\eea
Another exclusive decay which is described at inclusive level by the 
$b \rar s \ell^+ \ell^-$ transition is the baryonic 
$\Lambda_b \rar \Lambda \ell^+ \ell^-$ decay. Interest to the baryonic 
decays can be attributed to the fact that, unlike mesonic decays, they 
could maintain the helicity structure of the effective Hamiltonian for 
the $b \rar s$ transition. Note that, new physics effects in the 
$\Lambda_b \rar \Lambda \ell^+ \ell^-$ decay are studied in
\cite{R6520}.

Experimentally measurable quantities such as branching ratio \cite{R6521},
$\Lambda$ polarization and single lepton polarization have already been
studied in \cite{R6522} and \cite{R6523}, respectively. Study of such
quantities can give useful information for more precise determination of
the SM parameters and looking for new physics beyond the SM. It has been
pointed out in \cite{R6524} that the study of the polarizations of both
leptons provides measurement of many more observables which would be useful
in further improvement of the parameters of the SM and probing new physics
beyond the SM. 

In the present work we analyze the possibility of searching for new physics
in the baryonic $\Lambda_b \rar \Lambda \ell^+ \ell^-$ decay by studying the
polarized forward--backward asymmetry of various double--lepton
polarizations, using a general form of the effective Hamiltonian, including
all possible forms of interactions. It should be mentioned here that the
sensitivity of polarized forward--backward asymmetry to the new Wilson
coefficients for the meson$\rar$meson transition has been
investigated in \cite{R6525}. Naturally, one is forced to ask what happens
to this sensitivity in the case of baryon$\rar$baryon transition, i.e.,
which polarized asymmetry is sensitive to which new Wilson coefficients for
the baryon$\rar$baryon transition. A detailed investigation of this problem
is the main goal of the present work. 

The paper is organized as follows. In section
2, using the general, model independent form of the effective Hamiltonian,
the matrix element for the $\Lambda_b \rar \Lambda \ell^+ \ell^-$ is obtained.
In section 3 the analytic expressions for the polarized forward--backward
asymmetry are derived. Section 4 is devoted to the numerical analysis,
discussions and conclusions.

\section{Matrix element for the $\Lambda_b \rar \Lambda \ell^+ \ell^-$ decay}

In this section we derive the matrix element for the $\Lambda_b \rar \Lambda
\ell^+ \ell^-$ decay using the general, model independent form of the
effective Hamiltonian. At quark level, the matrix element of the 
$\Lambda_b \rar \Lambda \ell^+ \ell^-$ decay is described by the
$b \rar s \ell^+ \ell^-$ transition. The effective Hamiltonian for the
$b \rar s \ell^+ \ell^-$ transition can be written in terms of twelve model
independent four--Fermi interactions as \cite{R6515,R6526}:
\bea
\label{e6501}
{\cal M} \es \frac{G \alpha}{\sqrt{2} \pi} V_{tb}V_{ts}^\ast \Bigg\{
C_{SL} \bar s_R i \sigma_{\mu\nu} \frac{q^\nu}{q^2} b_L \bar \ell \gamma^\mu
\ell + C_{BR} \bar s_L i \sigma_{\mu\nu} \frac{q^\nu}{q^2} b_R \bar \ell
\gamma^\mu \ell + C_{LL}^{tot} \bar s_L \gamma_\mu b_L \bar \ell_L
\gamma^\mu \ell_L \nnb \\
\ar C_{LR}^{tot} \bar s_L \gamma_\mu b_L \bar \ell_R  
\gamma^\mu \ell_R + C_{RL} \bar s_R \gamma_\mu b_R \bar \ell_L
\gamma^\mu \ell_L + C_{RR} \bar s_R \gamma_\mu b_R \bar \ell_R
\gamma^\mu \ell_R \nnb \\
\ar C_{LRLR} \bar s_L b_R \bar \ell_L \ell_R +
C_{RLLR} \bar s_R b_L \bar \ell_L \ell_R +
C_{LRRL} \bar s_L b_R \bar \ell_R \ell_L +
C_{RLRL} \bar s_R b_L \bar \ell_R \ell_L \nnb \\
\ar C_T \bar s \sigma_{\mu\nu} b \bar \ell \sigma^{\mu\nu} \ell +
i C_{TE} \epsilon_{\mu\nu\alpha\beta} \bar s \sigma^{\mu\nu} b 
\bar \ell \sigma^{\alpha\beta} \ell \Bigg\}~,
\eea
where $L$ and $R$ are the chiral operators defined as
$L=(1-\gamma_5)/2$ and $R=(1+\gamma_5)/2$.
The coefficients of the first two terms, $C_{SL}$ and
$C_{BR}$ describe the penguin contributions, which correspond to 
$-2 m_s C_7^{eff}$ and $-2 m_b C_7^{eff}$ in the SM, respectively. 
The next four terms in Eq. (\ref{e6501}) with coefficients
$C_{LL}^{tot},~C_{LR}^{tot},~ C_{RL}$ and $C_{RR}$
describe vector type interactions. Two of these coefficients
$C_{LL}^{tot}$ and $C_{LR}^{tot}$ contain SM results in the form
$C_9^{eff}-C_{10}$ and $C_9^{eff}-C_{10}$, respectively.
Therefore, $C_{LL}^{tot}$ and $C_{LR}^{tot}$ can be written in the following
form: 
\bea
\label{e6502}
C_{LL}^{tot} \es C_9^{eff}- C_{10} + C_{LL}~, \nnb \\
C_{LR}^{tot} \es C_9^{eff}+ C_{10} + C_{LR}~,
\eea
where $C_{LL}$ and $C_{LR}$ describe the contributions of new physics. The
following four terms in Eq. (\ref{e6501}) with
coefficients $C_{LRLR},~C_{RLLR},~C_{LRRL}$ and $C_{RLRL}$ represent the
scalar type interactions. The remaining last two terms led by the
coefficients $C_T$ and $C_{TE}$ are the tensor type interactions.

The amplitude of the exclusive $\Lambda_b \rar \Lambda\ell^+ \ell^-$ decay
is obtained by calculating the matrix element of ${\cal H}_{eff}$ for the $b
\rar s \ell^+ \ell^-$ transition between initial and final
baryon states $\lla \Lambda \vel {\cal H}_{eff} \ver \Lambda_b \rra$.
We see from Eq. (\ref{e6501}) that for
calculating the $\Lambda_b \rar \Lambda\ell^+ \ell^-$ decay amplitude,
the following matrix elements are needed:

\bea
&&\lla \Lambda \vel \bar s \gamma_\mu (1 \mp \gamma_5) b \ver \Lambda_b
\rra~,\nnb \\
&&\lla \Lambda \vel \bar s \sigma_{\mu\nu} (1 \mp \gamma_5) b \ver \Lambda_b
\rra~,\nnb \\
&&\lla \Lambda \vel \bar s (1 \mp \gamma_5) b \ver \Lambda_b \rra~.\nnb
\eea

The relevant matrix elements parametrized in terms of the form factors are 
as follows (see \cite{R6522,R6527})
\bea
\label{e6503}
\lla \Lambda \vel \bar s \gamma_\mu b \ver \Lambda_b \rra  
\es \bar u_\Lambda \Big[ f_1 \gamma_\mu + i f_2 \sigma_{\mu\nu} q^\nu + f_3  
q_\mu \Big] u_{\Lambda_b}~,\\
\label{e6504}
\lla \Lambda \vel \bar s \gamma_\mu \gamma_5 b \ver \Lambda_b \rra
\es \bar u_\Lambda \Big[ g_1 \gamma_\mu \gamma_5 + i g_2 \sigma_{\mu\nu}
\gamma_5 q^\nu + g_3 q_\mu \gamma_5\Big] u_{\Lambda_b}~, \\
\label{e6505}
\lla \Lambda \vel \bar s \sigma_{\mu\nu} b \ver \Lambda_b \rra
\es \bar u_\Lambda \Big[ f_T \sigma_{\mu\nu} - i f_T^V \ga \gamma_\mu q^\nu -
\gamma_\nu q^\mu \dr - i f_T^S \ga P_\mu q^\nu - P_\nu q^\mu \dr \Big]
u_{\Lambda_b}~,\\
\label{e6506}
\lla \Lambda \vel \bar s \sigma_{\mu\nu} \gamma_5 b \ver \Lambda_b \rra
\es \bar u_\Lambda \Big[ g_T \sigma_{\mu\nu} - i g_T^V \ga \gamma_\mu q^\nu -
\gamma_\nu q^\mu \dr - i g_T^S \ga P_\mu q^\nu - P_\nu q^\mu \dr \Big]
\gamma_5 u_{\Lambda_b}~,
\eea
where $P = p_{\Lambda_b} + p_\Lambda$ and $q= p_{\Lambda_b} - p_\Lambda$. 

The form factors of the magnetic dipole operators are defined as 
\bea
\label{e6507}
\lla \Lambda \vel \bar s i \sigma_{\mu\nu} q^\nu  b \ver \Lambda_b \rra
\es \bar u_\Lambda \Big[ f_1^T \gamma_\mu + i f_2^T \sigma_{\mu\nu} q^\nu
+ f_3^T q_\mu \Big] u_{\Lambda_b}~,\nnb \\
\lla \Lambda \vel \bar s i \sigma_{\mu\nu}\gamma_5  q^\nu  b \ver \Lambda_b \rra
\es \bar u_\Lambda \Big[ g_1^T \gamma_\mu \gamma_5 + i g_2^T \sigma_{\mu\nu}
\gamma_5 q^\nu + g_3^T q_\mu \gamma_5\Big] u_{\Lambda_b}~.
\eea

Using the identity 
\bea
\sigma_{\mu\nu}\gamma_5 = - \frac{i}{2} \epsilon_{\mu\nu\alpha\beta}
\sigma^{\alpha\beta}~,\nnb
\eea
and Eq. (\ref{e6505}), the last expression in Eq. (\ref{e6507}) can be written as
\bea
\lla \Lambda \vel \bar s i \sigma_{\mu\nu}\gamma_5  q^\nu  b \ver \Lambda_b \rra
\es \bar u_\Lambda \Big[ f_T i \sigma_{\mu\nu} \gamma_5 q^\nu \Big]
u_{\Lambda_b}~.\nnb
\eea  
Multiplying (\ref{e6505}) and (\ref{e6506}) by $i q^\nu$ and comparing with
(\ref{e6507}), one can easily obtain the following relations between the form
factors
\bea
\label{e6508}
f_2^T \es f_T + f_T^S q^2~,\crcr
f_1^T \es \Big[ f_T^V + f_T^S \ga m_{\Lambda_b} + m_\Lambda\dr \Big] 
q^2~ = - \frac{q^2}{m_{\Lambda_b} - m_\Lambda} f_3^T~,\nnb \\
g_2^T \es g_T + g_T^S q^2~,\\
g_1^T \es \Big[ g_T^V - g_T^S \ga m_{\Lambda_b} - m_\Lambda\dr \Big]
q^2 =  \frac{q^2}{m_{\Lambda_b} + m_\Lambda} g_3^T~.\nnb
\eea 

The matrix element of the scalar $\bar s b$ and pseudoscalar
$\bar s\gamma_5 b$ operators can be obtained from (\ref{e6503}) 
and (\ref{e6504}) by multiplying both
sides to $q^\mu$ and using equation of motion. Neglecting the mass of the
strange quark, we get
\bea
\label{e6509}
\lla \Lambda \vel \bar s b \ver \Lambda_b \rra \es \frac{1}{m_b} 
\bar u_\Lambda \Big[ f_1 \ga m_{\Lambda_b} - m_\Lambda \dr + f_3 q^2
\Big] u_{\Lambda_b}~,\\
\label{e6510}
\lla \Lambda \vel \bar s \gamma_5 b \ver \Lambda_b \rra \es \frac{1}{m_b} 
\bar u_\Lambda \Big[ g_1 \ga m_{\Lambda_b} + m_\Lambda\dr \gamma_5 - g_3 q^2
\gamma_5 \Big] u_{\Lambda_b}~.
\eea

Using these definitions of the form factors, for the matrix element
of the $\Lambda_b \rar \Lambda\ell^+ \ell^-$ we get \cite{R6522}
\bea
\label{e6511}
\lefteqn{
{\cal M} = \frac{G \alpha}{4 \sqrt{2}\pi} V_{tb}V_{ts}^\ast \Bigg\{
\bar \ell \gamma^\mu \ell \, \bar u_\Lambda \Big[ A_1 \gamma_\mu (1+\gamma_5) +
B_1 \gamma_\mu (1-\gamma_5) }\nnb \\
\ar i \sigma_{\mu\nu} q^\nu \big[ A_2 (1+\gamma_5) + B_2 (1-\gamma_5) \big]
+q_\mu \big[ A_3 (1+\gamma_5) + B_3 (1-\gamma_5) \big]\Big] u_{\Lambda_b}
\nnb \\
\ar \bar \ell \gamma^\mu \gamma_5 \ell \, \bar u_\Lambda \Big[
D_1 \gamma_\mu (1+\gamma_5) + E_1 \gamma_\mu (1-\gamma_5) +
i \sigma_{\mu\nu} q^\nu \big[ D_2 (1+\gamma_5) + E_2 (1-\gamma_5) \big]
\nnb \\
\ar q_\mu \big[ D_3 (1+\gamma_5) + E_3 (1-\gamma_5) \big]\Big] u_{\Lambda_b}+
\bar \ell \ell\, \bar u_\Lambda \big(N_1 + H_1 \gamma_5\big) u_{\Lambda_b}
+\bar \ell \gamma_5 \ell \, \bar u_\Lambda \big(N_2 + H_2 \gamma_5\big) 
u_{\Lambda_b}\nnb \\
\ar 4 C_T \bar \ell \sigma^{\mu\nu}\ell \, \bar u_\Lambda \Big[ f_T 
\sigma_{\mu\nu} - i f_T^V \big( q_\nu \gamma_\mu - q_\mu \gamma_\nu \big) -
i f_T^S \big( P_\mu q_\nu - P_\nu q_\mu \big) \Big] u_{\Lambda_b}\nnb \\
\ar 4 C_{TE} \epsilon^{\mu\nu\alpha\beta} \bar \ell \sigma_{\alpha\beta}
\ell \, i \bar u_\Lambda \Big[ f_T \sigma_{\mu\nu} - 
i f_T^V \big( q_\nu \gamma_\mu - q_\mu \gamma_\nu \big) -
i f_T^S \big( P_\mu q_\nu - P_\nu q_\mu \big) \Big] u_{\Lambda_b}\Bigg\}~,
\eea
where the explicit forms of the functions $A_i,~B_i,~D_i,~E_i,~H_j$ and $N_j$
$(i=1,2,3$ and $j=1,2)$ can be written as \cite{R6522}

\bea
\label{e6512}
A_1 \es \frac{1}{q^2}\ga f_1^T-g_1^T \dr C_{SL} + \frac{1}{q^2}\ga
f_1^T+g_1^T \dr C_{BR} + \frac{1}{2}\ga f_1-g_1 \dr \ga C_{LL}^{tot} +
C_{LR}^{tot} \dr \nnb \\
\ar \frac{1}{2}\ga f_1+g_1 \dr \ga C_{RL} + C_{RR} \dr~,\nnb \\
A_2 \es A_1 \ga 1 \rar 2 \dr ~,\nnb \\
A_3 \es A_1 \ga 1 \rar 3 \dr ~,\nnb \\
B_1 \es A_1 \ga g_1 \rar - g_1;~g_1^T \rar - g_1^T \dr ~,\nnb \\
B_2 \es B_1 \ga 1 \rar 2 \dr ~,\nnb \\
B_3 \es B_1 \ga 1 \rar 3 \dr ~,\nnb \\
D_1 \es \frac{1}{2} \ga C_{RR} - C_{RL} \dr \ga f_1+g_1 \dr +
\frac{1}{2} \ga C_{LR}^{tot} - C_{LL}^{tot} \dr \ga f_1-g_1 \dr~,\nnb \\
D_2 \es D_1 \ga 1 \rar 2 \dr ~, \\
D_3 \es D_1 \ga 1 \rar 3 \dr ~,\nnb \\
E_1 \es D_1 \ga g_1 \rar - g_1 \dr ~,\nnb \\
E_2 \es E_1 \ga 1 \rar 2 \dr ~,\nnb \\
E_3 \es E_1 \ga 1 \rar 3 \dr ~,\nnb \\
N_1 \es \frac{1}{m_b} \Big( f_1 \ga m_{\Lambda_b} - m_\Lambda\dr + f_3 q^2
\Big) \Big( C_{LRLR} + C_{RLLR} + C_{LRRL} + C_{RLRL} \Big)~,\nnb \\
N_2 \es N_1 \ga C_{LRRL} \rar - C_{LRRL};~C_{RLRL} \rar - C_{RLRL} \dr~,\nnb \\
H_1 \es \frac{1}{m_b} \Big( g_1 \ga m_{\Lambda_b} + m_\Lambda\dr - g_3 q^2  
\Big) \Big( C_{LRLR} - C_{RLLR} + C_{LRRL} - C_{RLRL} \Big)~,\nnb \\
H_2 \es H_1 \ga C_{LRRL} \rar - C_{LRRL};~C_{RLRL} \rar - C_{RLRL} \dr~.\nnb
\eea

From these expressions it follows
that $\Lambda_b \rar\Lambda \ell^+\ell^-$ decay is described in terms of  
many form factors. It is shown in \cite{R6528} that when HQET is applied the
number of independent form factors reduces to two ($F_1$ and
$F_2$) irrelevant of the Dirac structure
of the corresponding operators, i.e., 
\bea
\label{e6513}
\lla \Lambda(p_\Lambda) \vel \bar s \Gamma b \ver \Lambda(p_{\Lambda_b})
\rra = \bar u_\Lambda \Big[F_1(q^2) + \not\!v F_2(q^2)\Big] \Gamma
u_{\Lambda_b}~,
\eea
where $\Gamma$ is an arbitrary Dirac structure and
$v^\mu=p_{\Lambda_b}^\mu/m_{\Lambda_b}$ is the four--velocity of
$\Lambda_b$. Comparing the general form of the form factors given in Eqs.
(\ref{e6503})--(\ref{e6510}) with (\ref{e6513}), one can
easily obtain the following relations among them (see also \cite{R6527})
\bea
\label{e6514}
g_1 \es f_1 = f_2^T= g_2^T = F_1 + \sqrt{\hat{r}_\Lambda} F_2~, \nnb \\
g_2 \es f_2 = g_3 = f_3 = g_T^V = f_T^V = \frac{F_2}{m_{\Lambda_b}}~,\nnb \\
g_T^S \es f_T^S = 0 ~,\nnb \\
g_1^T \es f_1^T = \frac{F_2}{m_{\Lambda_b}} q^2~,\nnb \\
g_3^T \es \frac{F_2}{m_{\Lambda_b}} \ga m_{\Lambda_b} + m_\Lambda \dr~,\nnb \\
f_3^T \es - \frac{F_2}{m_{\Lambda_b}} \ga m_{\Lambda_b} - m_\Lambda \dr~,
\eea
where $\hat{r}_\Lambda=m_\Lambda^2/m_{\Lambda_b}^2$.

From Eq. (\ref{e6511}), we get for the unpolarized decay width
\bea
\label{e6515}
\ga \frac{d \Gamma}{d\hat{s}}\dr_0 = \frac{G^2 \alpha^2}{8192 \pi^5}
\vel V_{tb} V_{ts}^\ast \ver^2 \lambda^{1/2}(1,\hat{r}_\Lambda,\hat{s}) v
\Bigg[{\cal T}_0(\hat{s}) +\frac{1}{3} {\cal T}_2(\hat{s}) \Bigg]~,
\eea
where 
$\lambda(1,\hat{r}_\Lambda,\hat{s}) = 1 + \hat{r}_\Lambda^2 + \hat{s}^2 - 
2 \hat{r}_\Lambda - 2 \hat{s} - 2 \hat{r}_\Lambda\hat{s}$
is the triangle function, $\hat{s} = q^2/m_{\Lambda_b}^2$ and
$v=\sqrt{1-4 \hat{m}_\ell^2/\hat{s}}$ is the lepton
velocity, with $\hat{m}_\ell = m_\ell/m_{\Lambda_b}$. 
The explicit expressions for ${\cal T}_0$ and ${\cal T}_2$ can be 
found in \cite{R6522}. The expressions for the doubly--polarized lepton pair
forward--backward asymmetry will be presented in the next section.

\section{Polarized forward--backward asymmetries of leptons in the 
$\Lambda_b \rar\Lambda \ell^+\ell^-$ decay}

In this section we calculate the polarized forward--backward asymmetries,
and for this aim we define the following orthogonal unit vectors
$s_i^{\pm\mu}$ in the rest frame of $\ell^\pm$ 
($i=L,T$ or $N$, stand for longitudinal, transversal or
normal polarizations, respectively. See also \cite{R6515},
\cite{R6529}--\cite{R6531})

\bea
\label{e6516}   
s^{-\mu}_L \es \ga 0,\vec{e}_L^{\,-}\dr =
\ga 0,\frac{\vec{p}_-}{\vel\vec{p}_- \ver}\dr~, \nnb \\
s^{-\mu}_N \es \ga 0,\vec{e}_N^{\,-}\dr = \ga 0,\frac{\vec{p}_\Lambda\times
\vec{p}_-}{\vel \vec{p}_\Lambda\times \vec{p}_- \ver}\dr~, \nnb \\
s^{-\mu}_T \es \ga 0,\vec{e}_T^{\,-}\dr = \ga 0,\vec{e}_N^{\,-}
\times \vec{e}_L^{\,-} \dr~, \nnb \\
s^{+\mu}_L \es \ga 0,\vec{e}_L^{\,+}\dr =
\ga 0,\frac{\vec{p}_+}{\vel\vec{p}_+ \ver}\dr~, \nnb \\
s^{+\mu}_N \es \ga 0,\vec{e}_N^{\,+}\dr = \ga 0,\frac{\vec{p}_\Lambda\times
\vec{p}_+}{\vel \vec{p}_\Lambda\times \vec{p}_+ \ver}\dr~, \nnb \\
s^{+\mu}_T \es \ga 0,\vec{e}_T^{\,+}\dr = \ga 0,\vec{e}_N^{\,+}
\times \vec{e}_L^{\,+}\dr~,
\eea
where $\vec{p}_\mp$ and $\vec{p}_\Lambda$ are the three--momenta of the
leptons $\ell^\mp$ and $\Lambda$ baryon in the
center of mass frame (CM) of $\ell^- \,\ell^+$ system, respectively.
Transformation of unit vectors from the rest frame of the leptons to CM
frame of leptons can be accomplished by the Lorentz boost. Boosting of the
longitudinal unit vectors $s_L^{\pm\mu}$ yields
\bea
\label{e6517}
\ga s^{\mp\mu}_L \dr_{CM} \es \ga \frac{\vel\vec{p}_\mp \ver}{m_\ell}~,
\frac{E_\ell \vec{p}_\mp}{m_\ell \vel\vec{p}_\mp \ver}\dr~,
\eea
where $\vec{p}_+ = - \vec{p}_-$, $E_\ell$ and $m_\ell$ are the energy and mass
of leptons in the CM frame, respectively.
The remaining two unit vectors $s_N^{\pm\mu}$, $s_T^{\pm\mu}$ are unchanged
under Lorentz boost.

The definition of the normalized, unpolarized differential
forward--backward asymmetry is
\bea
\label{e6518}
{\cal A}_{FB} = \frac{\ds \int_{0}^{1} \frac{d^2\Gamma}{d\hat{s} dz} -
\int_{-1}^{0} \frac{d^2\Gamma}{d\hat{s} dz}}
{\ds \int_{0}^{1} \frac{d^2\Gamma}{d\hat{s} dz} +
\int_{-1}^{0} \frac{d^2\Gamma}{d\hat{s} dz}}~,
\eea
where $z=\cos\theta$ is the angle between $\Lambda_b$ meson and $\ell^-$ in the
center mass frame of leptons. When the spins of both leptons are taken into
account, the ${\cal A}_{FB}$ will be a function of the spins of the final
leptons and it is defined as
\bea
\label{e6519}
{\cal A}_{FB}^{ij}(\hat{s}) \es 
\Bigg(\frac{d\Gamma(\hat{s})}{d\hat{s}} \Bigg)^{-1}
\Bigg\{ \int_0^1 dz - \int_{-1}^0 dz \Bigg\}
\Bigg\{ 
\Bigg[
\frac{d^2\Gamma(\hat{s},\vec{s}^{\,-} = \vec{i},\vec{s}^{\,+} = \vec{j})}
{d\hat{s} dz} - 
\frac{d^2\Gamma(\hat{s},\vec{s}^{\,-} = \vec{i},\vec{s}^{\,+} = -\vec{j})} 
{d\hat{s} dz}
\Bigg] \nnb \\
\ek
\Bigg[
\frac{d^2\Gamma(\hat{s},\vec{s}^{\,-} = -\vec{i},\vec{s}^{\,+} = \vec{j})} 
{d\hat{s} dz} - 
\frac{d^2\Gamma(\hat{s},\vec{s}^{\,-} = -\vec{i},\vec{s}^{\,+} = -\vec{j})} 
{d\hat{s} dz}
\Bigg]
\Bigg\}~,\nnb \\ \nnb \\
\es 
{\cal A}_{FB}(\vec{s}^{\,-}=\vec{i},\vec{s}^{\,+}=\vec{j})   -
{\cal A}_{FB}(\vec{s}^{\,-}=\vec{i},\vec{s}^{\,+}=-\vec{j})  - 
{\cal A}_{FB}(\vec{s}^{\,-}=-\vec{i},\vec{s}^{\,+}=\vec{j})  \nnb \\
\ar   
{\cal A}_{FB}(\vec{s}^{\,-}=-\vec{i},\vec{s}^{\,+}=-\vec{j})~.   
\eea

Using these definitions for the double polarized $FB$ asymmetries, we get
the following results:   

\bea
\label{e6520}
{\cal A}_{FB}^{LL} \es \frac{16}{\Delta} m_{\Lambda_b}^4 \sqrt{\lambda} v 
\mbox{\rm Re} \Big[
- \hat{m}_\ell  \Big(1-\sqrt{\hat{r}_\Lambda}\Big) 
(A_1-B_1) H_1^\ast \nnb \\
\ar \hat{m}_\ell \Big(1+\sqrt{\hat{r}_\Lambda}\Big)
(A_1+B_1) F_1^\ast  \nnb \\
\ar 8 \hat{m}_\ell \Big\{ 
2 \Big(1-\sqrt{\hat{r}_\Lambda}\Big) 
(D_1+E_1) f_T^\ast C_{TE}^\ast +
\Big(1+\sqrt{\hat{r}_\Lambda}\Big) 
(D_1-E_1) f_T^\ast C_{T}^\ast \Big\} \nnb \\
\ar 16 m_{\Lambda_b} \hat{m}_\ell (1-\hat{r}_\Lambda )\Big\{
(D_2 - E_2) f_T^\ast C_{T}^\ast - 2 (D_2 + E_2) f_T^\ast  
C_{TE}^\ast \Big\} \nnb \\
\ar 16 m_{\Lambda_b}^2 \hat{m}_\ell 
\Big(1-\sqrt{\hat{r}_\Lambda} \Big) \Big(1+2 \sqrt{\hat{r}_\Lambda}
+ \hat{r}_\Lambda -\hat{s}\Big) 
(D_1 + E_1) f_T^{S\ast} C_{TE}^\ast \nnb \\
\ar 2 \hat{s}
\Big\{ A_1 D_1^\ast - B_1 E_1^\ast 
- 4 (F_2+H_1) f_T^\ast C_{TE}^\ast + 2 (F_1+H_2) f_T^\ast C_{T}^\ast
\Big\} \nnb \\
\ek m_{\Lambda_b} \hat{s}
\Big\{ 2(A_1 E_2^\ast - A_2 E_1^\ast - B_1 D_2^\ast +    
B_2 D_1^\ast) + \hat{m}_\ell (A_2 + B_2 ) F_1^\ast \Big\} \nnb \\
\ar 8 m_{\Lambda_b} \hat{m}_\ell \hat{s} \Big\{
(D_3 - E_3) f_T^\ast C_{T}^\ast - 2 (D_3 + E_3) f_T^\ast  
C_{TE}^\ast \Big\}\nnb \\
\ek m_{\Lambda_b} \hat{m}_\ell \hat{s}
(A_2 - B_2) H_1^\ast \nnb \\
\ek 16 m_{\Lambda_b}^2 \hat{m}_\ell \hat{s}
\Big(1-\sqrt{\hat{r}_\Lambda} \Big)
(D_2 - E_2) f_T^{V\ast} C_{T}^\ast \nnb \\
\ek 4 m_{\Lambda_b} \hat{s}
\Big(1+\sqrt{\hat{r}_\Lambda} \Big)
\Big( F_1 f_T^{V\ast} C_{T}^\ast - 
2 F_2 f_T^{V\ast} C_{TE}^\ast \Big) \nnb \\
\ar 16 m_{\Lambda_b}^2 \hat{m}_\ell \hat{s}
\Big(1+\sqrt{\hat{r}_\Lambda} \Big)
(D_3 + E_3) f_T^{V\ast} C_{TE}^\ast \nnb \\
\ek 2 m_{\Lambda_b} \hat{s}
\Big\{  m_{\Lambda_b} (1-\hat{r}_\Lambda ) (A_2 D_2^\ast -
B_2 E_2^\ast) + \sqrt{\hat{r}_\Lambda} (A_1 D_2^\ast + A_2 D_1^\ast) 
\Big\} \nnb \\
\ar 2 m_{\Lambda_b} \hat{s} \sqrt{\hat{r}_\Lambda}
( B_1 E_2^\ast + B_2 E_1^\ast ) \nnb \\
\ar 16 m_{\Lambda_b}^3 \hat{m}_\ell \hat{s} 
\Big( 1 + 2 \sqrt{\hat{r}_\Lambda} + \hat{r}_\Lambda - \hat{s} \Big)
(D_3 + E_3) f_T^{S\ast} C_{TE}^\ast \nnb \\
\ek 4 m_{\Lambda_b}^2 \hat{s}
\Big( 1 + 2 \sqrt{\hat{r}_\Lambda} + \hat{r}_\Lambda - \hat{s} \Big)
\Big( F_1 f_T^{S\ast} C_{T}^\ast - 
2 F_2 f_T^{S\ast} C_{TE}^\ast \Big) \nnb \\
\ar 16 m_{\Lambda_b} \hat{m}_\ell
\Big\{ (1-\hat{r}_\Lambda)      
(D_1 + E_1) f_T^{V\ast} C_{TE}^\ast -
\hat{s} (D_1 - E_1) f_T^{V\ast} C_{T}^\ast \Big\} 
\Big]~, \\ \nnb \\
\label{e6521}
{\cal A}_{FB}^{LT} \es \frac{64}{3 \sqrt{\hat{s}}\Delta} m_{\Lambda_b}^4 \lambda
\mbox{\rm Re} \Big[
- \hat{m}_\ell
\Big\{ \vel A_1 \ver^2 + \vel B_1 \ver^2 - 32 \Big( 4 \vel C_{TE} \ver^2 +
\vel C_{T} \ver^2 \Big) \vel f_T \ver^2 \Big\} \nnb \\
\ar m_{\Lambda_b}^2 \hat{m}_\ell \hat{s}  
\Big\{ \vel A_2 \ver^2 + \vel B_2 \ver^2 + 32 \vel C_{T} \ver^2 \Big( 
2 f_T f_T^{S\ast} - \vel f_T^V \ver^2 \Big) 
\Big\} \nnb \\
\ek 8 m_{\Lambda_b} \hat{s} v^2 
\Big\{ (A_2 + B_2) f_T^\ast C_{T}^\ast -
2 (A_2 - B_2) f_T^\ast C_{TE}^\ast - A_1 f_T^{V\ast} C_{T}^\ast 
+ m_{\Lambda_b}^2 \hat{s} (A_2 + B_2) f_T^{S\ast} C_{T}^\ast\Big\} \nnb \\
\ar 8 m_{\Lambda_b}  \hat{s} v^2
\Big\{ B_1 f_T^{V\ast} C_T^\ast + m_{\Lambda_b} 
\Big(1 + \sqrt{\hat{r}_{\Lambda}} \Big) A_1 f_T^{S\ast} C_T^\ast 
\Big\} \nnb \\
\ar 8 m_{\Lambda_b}^2 \hat{s}    
\Big(1 + \sqrt{\hat{r}_{\Lambda}} \Big)
\Big( - 32 m_{\Lambda_b} \hat{m}_\ell \vel C_{T} \ver^2
f_T^S f_T^{V\ast} + v^2 B_1 C_{T}^\ast f_T^{S\ast} \Big) \nnb \\
\ar 8 m_{\Lambda_b} \hat{s} v^2 
\Big\{ (D_2 - E_2) f_T^\ast C_{T}^\ast -  
2 (D_2 + E_2) f_T^\ast C_{TE}^\ast +  
2 (D_1 + E_1) f_T^{V\ast} C_{TE}^\ast \Big\} \nnb \\
\ar 16 m_{\Lambda_b}^2 \hat{s} v^2 
\Big(1 + \sqrt{\hat{r}_{\Lambda}} \Big)
(D_1 + E_1) f_T^{S\ast} C_{TE}^\ast \nnb \\
\ek 16 m_{\Lambda_b}^3 \hat{s}^2 v^2
(D_2 + E_2) f_T^{S\ast} C_{TE}^\ast \nnb \\
\ek 128 m_{\Lambda_b}^4 \hat{m}_\ell \hat{s}    
\Big( 1 + 2 \sqrt{\hat{r}_\Lambda} + \hat{r}_\Lambda - \hat{s} \Big)
\vel C_{T} \ver^2 \vel f_T^{S\ast} \ver^2 
\Big]~, \\ \nnb \\
\label{e6522}
{\cal A}_{FB}^{TL} \es \frac{64}{3 \sqrt{\hat{s}}\Delta} m_{\Lambda_b}^4 \lambda
\mbox{\rm Re} \Big[
\hat{m}_\ell          
\Big\{ \vel A_1 \ver^2 + \vel B_1 \ver^2 - 32 \Big( 4 \vel C_{TE} \ver^2 +
\vel C_{T} \ver^2 \Big) \vel f_T \ver^2 \Big\} \nnb \\
\ek m_{\Lambda_b}^2 \hat{m}_\ell \hat{s}  
\Big\{ \vel A_2 \ver^2 + \vel B_2 \ver^2 + 32 \vel C_{T} \ver^2 \Big( 
2 f_T f_T^{S\ast} - \vel f_T^V \ver^2 \Big) 
\Big\} \nnb \\
\ar 8 m_{\Lambda_b} \hat{s} v^2 
\Big\{ (A_2 + B_2) f_T^\ast C_{T}^\ast -
2 (A_2 - B_2) f_T^\ast C_{TE}^\ast - A_1 f_T^{V\ast} C_{T}^\ast 
+ m_{\Lambda_b}^2 \hat{s} (A_2 + B_2) f_T^{S\ast} C_{T}^\ast\Big\} \nnb \\
\ek 8 m_{\Lambda_b} \hat{s} v^2
\Big\{ B_1 f_T^{V\ast} C_T^\ast + m_{\Lambda_b} 
\Big(1 + \sqrt{\hat{r}_{\Lambda}} \Big) A_1 f_T^{S\ast} C_T^\ast 
\Big\} \nnb \\
\ek 8 m_{\Lambda_b}^2 \hat{s}    
\Big(1 + \sqrt{\hat{r}_{\Lambda}} \Big)
\Big( - 32 m_{\Lambda_b} \hat{m}_\ell \vel C_{T} \ver^2
f_T^S f_T^{V\ast} + v^2 B_1 C_{T}^\ast f_T^{S\ast} \Big) \nnb \\
\ar 8 m_{\Lambda_b} \hat{s} v^2 
\Big\{ (D_2 - E_2) f_T^\ast C_{T}^\ast -  
2 (D_2 + E_2) f_T^\ast C_{TE}^\ast +  
2 (D_1 + E_1) f_T^{V\ast} C_{TE}^\ast \Big\} \nnb \\
\ar 16 m_{\Lambda_b}^2 \hat{s} v^2 
\Big(1 + \sqrt{\hat{r}_{\Lambda}} \Big)
(D_1 + E_1) f_T^{S\ast} C_{TE}^\ast \nnb \\
\ek 16 m_{\Lambda_b}^3 \hat{s}^2 v^2
(D_2 + E_2) f_T^{S\ast} C_{TE}^\ast \nnb \\
\ar 128 m_{\Lambda_b}^4 \hat{m}_\ell \hat{s}    
\Big( 1 + 2 \sqrt{\hat{r}_\Lambda} + \hat{r}_\Lambda - \hat{s} \Big)
\vel C_{T} \ver^2 \vel f_T^{S\ast} \ver^2 
\Big]~, \\ \nnb \\
\label{e6523}
{\cal A}_{FB}^{LN} \es \frac{64}{3 \sqrt{\hat{s}}\Delta} m_{\Lambda_b}^4 \lambda v
\mbox{\rm Im} \Big[
- \hat{m}_\ell
( A_1 D_1^\ast + B_1 E_1^\ast ) \nnb \\
\ek 2 m_{\Lambda_b} \hat{s}
\Big\{ (A_2 - D_2) f_T^\ast (C_T^\ast - 2 C_{TE}^\ast)
- (B_2 + E_2) f_T^\ast (C_T^\ast + 2 C_{TE}^\ast) \Big\} \nnb \\
\ek 2 m_{\Lambda_b} \hat{s}
\Big\{ 2 (A_1 + B_1) f_T^{V\ast} C_{TE}^\ast +
(D_1 + E_1) f_T^{V\ast} C_{T}^\ast \Big\} \nnb \\
\ar m_{\Lambda_b}^2 \hat{m}_\ell \hat{s}
( A_2 D_2^\ast + B_2 E_2^\ast ) \nnb \\
\ek 64 m_{\Lambda_b}^2 \hat{m}_\ell \hat{s}
\Big(2 \mbox{\rm Re} \Big[f_T f_T^{S\ast}\Big] - \vel f_T^V \ver^2 \Big)
C_{T} C_{TE}^\ast \nnb \\
\ek 2 m_{\Lambda_b}^2 \hat{s}
\Big( 1 + \sqrt{\hat{r}_\Lambda} \Big)
\Big\{ 2 (A_1 + B_1) f_T^{S\ast} C_{TE}^\ast +
(D_1 + E_1) f_T^{S\ast} C_{T}^\ast \Big\} \nnb \\
\ar 2 m_{\Lambda_b}^3 \hat{s}^2
\Big\{ 2 (A_2 + B_2) f_T^{S\ast} C_{TE}^\ast +
(D_2 + E_2) f_T^{S\ast} C_{T}^\ast \Big\} \nnb \\
\ar 64 m_{\Lambda_b}^3 \hat{m}_\ell \hat{s}
\Big\{ 2 \Big( 1 + \sqrt{\hat{r}_\Lambda} \Big) 
\mbox{\rm Re} \Big[f_T^S f_T^{V\ast}\Big] +
m_{\Lambda_b} \Big( 1 + 2 \sqrt{\hat{r}_\Lambda} +
\hat{r}_\Lambda - \hat{s} \Big)
 \vel f_T^S \ver^2 \Big\}
C_{T} C_{TE}^\ast 
\Big]~, \\ \nnb \\
\label{e6524}
{\cal A}_{FB}^{NL} \es \frac{64}{3 \sqrt{\hat{s}}\Delta} m_{\Lambda_b}^4 \lambda v
\mbox{\rm Im} \Big[
- \hat{m}_\ell 
( A_1 D_1^\ast + B_1 E_1^\ast ) \nnb \\
\ar 2 m_{\Lambda_b} \hat{s}
\Big\{ (A_2 + D_2) f_T^\ast (C_T^\ast - 2 C_{TE}^\ast)
- (B_2 - E_2) f_T^\ast (C_T^\ast + 2 C_{TE}^\ast) \Big\} \nnb \\
\ar 2 m_{\Lambda_b} \hat{s}
\Big\{ 2 (A_1 + B_1) f_T^{V\ast} C_{TE}^\ast -
(D_1 + E_1) f_T^{V\ast} C_{T}^\ast \Big\} \nnb \\
\ar m_{\Lambda_b}^2 \hat{m}_\ell \hat{s}
( A_2 D_2^\ast + B_2 E_2^\ast ) \nnb \\
\ar 64 m_{\Lambda_b}^2 \hat{m}_\ell \hat{s}
\Big(2 \mbox{\rm Re} \Big[f_T f_T^{S\ast}\Big] - \vel f_T^V \ver^2 \Big)
C_{T} C_{TE}^\ast \nnb \\
\ar 2 m_{\Lambda_b}^2 \hat{s}
\Big( 1 + \sqrt{\hat{r}_\Lambda} \Big)
\Big\{ 2 (A_1 + B_1) f_T^{S\ast} C_{TE}^\ast -
(D_1 + E_1) f_T^{S\ast} C_{T}^\ast \Big\} \nnb \\
\ek 2 m_{\Lambda_b}^3 \hat{s}^2
\Big\{ 2 (A_2 + B_2) f_T^{S\ast} C_{TE}^\ast -
(D_2 + E_2) f_T^{S\ast} C_{T}^\ast \Big\} \nnb \\
\ek 64 m_{\Lambda_b}^3 \hat{m}_\ell \hat{s}
\Big\{ 2 \Big( 1 + \sqrt{\hat{r}_\Lambda} \Big) 
\mbox{\rm Re} \Big[f_T^S f_T^{V\ast}\Big] +
m_{\Lambda_b} \Big( 1 + 2 \sqrt{\hat{r}_\Lambda} +
\hat{r}_\Lambda - \hat{s} \Big)
 \vel f_T^S \ver^2 \Big\}
C_{T} C_{TE}^\ast 
\Big]~, \\ \nnb \\ 
\label{e6525}
{\cal A}_{FB}^{NT} \es {\cal A}_{FB}^{TN} \nnb \\
\es \frac{16}{\Delta} m_{\Lambda_b}^4 \sqrt{\lambda}
\mbox{\rm Im} \Big[
4 m_{\Lambda_b} \hat{m}_\ell^2 
\Big\{ A_1 E_3^\ast - A_2 E_1^\ast + B_1 D_3^\ast 
- B_2 D_1^\ast \Big\} \nnb \\
\ek \hat{m}_\ell \Big( 1 - \sqrt{\hat{r}_\Lambda} \Big) 
(A_1 -B_1) H_2^\ast \nnb \\
\ar \hat{m}_\ell \Big( 1 + \sqrt{\hat{r}_\Lambda} \Big)
(A_1 +B_1) F_2^\ast \nnb \\
\ek 8 \hat{m}_\ell
\Big\{ \Big( 1 - \sqrt{\hat{r}_\Lambda} \Big)
(D_1 + E_1) f_T^\ast C_{T}^\ast +
2 \Big( 1 + \sqrt{\hat{r}_\Lambda} \Big) 
(D_1 - E_1) f_T^\ast C_{TE}^\ast \Big\} \nnb \\
\ar 8 m_{\Lambda_b} \hat{m}_\ell 
( 1 - \hat{r}_\Lambda )
(D_1 + E_1) f_T^{V^\ast} C_{T}^\ast \nnb \\
\ar 4 m_{\Lambda_b} \hat{m}_\ell^2 \sqrt{\hat{r}_\Lambda}
( A_1 D_3^\ast + A_2 D_1^\ast + B_1 E_3^\ast +B_2 E_1^\ast ) 
\nnb \\
\ar 8 m_{\Lambda_b}^2 \hat{m}_\ell
\Big( 1 - \sqrt{\hat{r}_\Lambda} \Big)
\Big( 1 + 2 \sqrt{\hat{r}_\Lambda} + \hat{r}_\Lambda - \hat{s}\Big)
(D_1 + E_1) f_T^{S\ast} C_{T}^\ast \nnb \\
\ar \frac{4}{\hat{s}} \hat{m}_\ell^2 
( 1 - \hat{r}_\Lambda)
( A_1 D_1^\ast + B_1 E_1^\ast ) \nnb \\
\ek m_{\Lambda_b} \hat{m}_\ell \hat{s} 
\Big\{ (A_2 + B_2) F_2^\ast + 
(A_2 - B_2) H_2^\ast \Big\} \nnb \\
\ar 4 \hat{s}
\Big[ 2 \{ H_2 + 2 m_{\Lambda_b} \hat{m}_\ell 
(D_3 - E_3) \} f_T^\ast C_{TE}^\ast - 
 \{ F_2 + 2 m_{\Lambda_b} \hat{m}_\ell 
(D_3 + E_3) \} f_T^\ast C_{T}^\ast \Big] \nnb \\
\ek 4 m_{\Lambda_b}^2 \hat{m}_\ell^2 \hat{s}
( A_2 D_3^\ast + B_2 E_3^\ast ) \nnb \\
\ar 4 m_{\Lambda_b} \hat{s} 
\Big( 1 + \sqrt{\hat{r}_\Lambda} \Big)
\{ F_2 + 2  m_{\Lambda_b} \hat{m}_\ell (D_3 + E_3) \}
f_T^{V\ast} C_{T}^\ast \nnb \\
\ar 4 m_{\Lambda_b}^2 \hat{s}
\Big( 1 + 2 \sqrt{\hat{r}_\Lambda} + \hat{r}_\Lambda - \hat{s}\Big)
\{ F_2 + 2  m_{\Lambda_b} \hat{m}_\ell (D_3 + E_3) \}
f_T^{S\ast} C_{T}^\ast \nnb \\
\ek 4 \hat{s} v^2 
\Big( 2 F_1 f_T^\ast C_{TE}^\ast - H_1 f_T^\ast C_{T}^\ast   
\Big) \nnb \\
\ar 8 m_{\Lambda_b} \hat{s} v^2
\Big\{ \Big( 1 + \sqrt{\hat{r}_\Lambda} \Big)
F_1 f_T^{V\ast} C_{TE}^\ast +
m_{\Lambda_b} \Big( 1 + 2 \sqrt{\hat{r}_\Lambda} + \hat{r}_\Lambda -
\hat{s}\Big)
F_1 f_T^{S\ast} C_{TE}^\ast \Big\} 
\Big]~, \\ \nnb \\
\label{e6526}
{\cal A}_{FB}^{NN} \es - {\cal A}_{FB}^{TT} \nnb \\
\es \frac{16}{\Delta} m_{\Lambda_b}^4 \sqrt{\lambda} v
\mbox{\rm Re} \Big[
- \hat{m}_\ell
\Big\{ (A_1 + B_1) F_1^\ast - (A_1 - B_1) H_1^\ast
+ \sqrt{\hat{r}_\Lambda} \{ (A_1 + B_1) F_1^\ast + (A_1 - B_1) H_1^\ast \}
\Big\} \nnb \\
\ek 8 \hat{m}_\ell
\Big\{ \Big( 1 + \sqrt{\hat{r}_\Lambda}\Big)
(D_1 - E_1) f_T^\ast C_{T}^\ast +   
2 \Big( 1 - \sqrt{\hat{r}_\Lambda}\Big)
(D_1 + E_1) f_T^\ast C_{TE}^\ast \Big\} \nnb \\
\ar 16 m_{\Lambda_b} \hat{m}_\ell (1 - \hat{r}_\Lambda)
(D_1 + E_1) f_T^{V\ast} C_{TE}^\ast \nnb \\
\ar 16 m_{\Lambda_b}^2 \hat{m}_\ell
\Big( 1 - \sqrt{\hat{r}_\Lambda} \Big)
\Big( 1 + 2 \sqrt{\hat{r}_\Lambda} + \hat{r}_\Lambda -\hat{s} \Big)
(D_1 + E_1) f_T^{S\ast} C_{TE}^\ast \nnb \\
\ek 4 \hat{s}
\Big\{ (F_1 - H_2) f_T^\ast C_{T}^\ast +
2 (F_2 - H_1) f_T^\ast C_{TE}^\ast \Big\} \nnb \\
\ar m_{\Lambda_b} \hat{m}_\ell \hat{s}
\Big\{ (A_2 + B_2) F_1^\ast + (A_2 - B_2) H_1^\ast \Big\} \nnb \\
\ar 8 m_{\Lambda_b} \hat{m}_\ell \hat{s}
\Big\{ (D_3 - E_3) f_T^\ast C_{T}^\ast
- 2 (D_3 + E_3) f_T^\ast C_{TE}^\ast \Big\} \nnb \\
\ar 4 m_{\Lambda_b} \hat{s} 
\Big( 1 + \sqrt{\hat{r}_\Lambda} \Big)
\Big( F_1 f_T^{V\ast} C_{T}^\ast + 
2 \{ F_2 + 2 m_{\Lambda_b} \hat{m}_\ell (D_3 + E_3) \} f_T^{V\ast}
C_{TE}^\ast \Big) \nnb \\
\ar 4 m_{\Lambda_b}^2 \hat{s}
\Big( 1 + 2 \sqrt{\hat{r}_\Lambda} + \hat{r}_\Lambda - \hat{s} \Big)
\Big( F_1 f_T^{S\ast} C_{T}^\ast +                   
2 \{ F_2 + 2 m_{\Lambda_b} \hat{m}_\ell (D_3 + E_3) \} f_T^{S\ast}
C_{TE}^\ast \Big) 
\Big]~.
\eea

In the expressions for ${\cal A}_{FB}^{ij}$, the superscript indices
$i$ and $j$ correspond to the lepton and anti--lepton polarizations, 
respectively.

At the end of this section, we would like to remind the interested reader
that, the doubly polarized forward--backward asymmetry is calculated in a
model independent way for the $B \rar K^\ast \ell^+ \ell^-$ decay, in SUSY
theories for the $B \rar K^\ast \tau^+ \tau^-$ decay and $b \rar s \tau^+
\tau^-$ transition in \cite{R6525}, \cite{R6532} and \cite{R6533}, 
respectively.

\section{Numerical analysis}

In this section we study the influence of the new Wilson coefficients on the
polarized forward--backward asymmetry. Before performing numerical
calculations, we present the values of the input parameters.$\vel V_{tb}
V_{ts}^\ast \ver = 0.0385$, $m_\tau = 1.77~GeV$, $m_\mu = 0.106~GeV$.
$m_b = 4.8~GeV$. For the values of the Wilson coefficients in SM we use
$C_7^{SM} = -0.313$, $C_9^{SM} = 4.344$ and $C_{10}^{SM} = -4.669$. The
value of $C_9^{SM}$ we use corresponds to the short distance contributions.
It is well known that, in addition to the short distance contributions,
$C_9$ receives long distance contributions, coming from the production of
$\bar{c}c$ pair at intermediate states. In the present work we neglect such
long distance effects. From the expressions of ${\cal A}_{FB}^{ij}$, it can
easily be seen that, one of the most important input parameters necessary in
the numerical calculations are the form factors. So far, the
calculations for all of the form factors of the $\Lambda_b \rar \Lambda$
transition are absent. Therefore, we will use the results from QCD sum rules
in corporation with HQET \cite{R6528,R6534}. As has already been noted,
HQET allows us to establish relations among the form factors and reduces
the number of independent form factors into two. 
In \cite{R6528,R6534}, the $q^2$ dependence of these form factors
are given as follows
\bea
\label{e19}
F(\hat{s}) = \frac{F(0)}{\ds 1-a_F \hat{s} + b_F \hat{s}^2}~. \nnb
\eea
The values of the parameters $F(0),~a_F$ and $b_F$ are given in table 1.
\begin{table}[h]    
\renewcommand{\arraystretch}{1.5}
\addtolength{\arraycolsep}{3pt}
$$
\begin{array}{|l|ccc|}  
\hline
& F(0) & a_F & b_F \\ \hline
F_1 &
\phantom{-}0.462 & -0.0182 & -0.000176 \\
F_2 &
-0.077 & -0.0685 &\phantom{-}0.00146 \\ \hline
\end{array}
$$
\caption{Form factors for $\Lambda_b \rar \Lambda \ell^+ \ell^-$
decay in a three parameter fit.}
\renewcommand{\arraystretch}{1}
\addtolength{\arraycolsep}{-3pt}
\end{table}  

In further numerical analysis, the values of the new Wilson coefficients
which describe new physics beyond the SM, are needed. In numerical
calculations we will vary all new Wilson coefficients in the range
$- \vel C_{10}^{SM} \ver \le C_X \le \vel C_{10}^{SM} \ver$. The
experimental results on the branching ratio of the 
$B \rar K^\ast \ell^+ \ell^-$ decay \cite{R6518,R6519} and the bound on the
branching ratio of $B \rar \mu^+ \mu^-$ \cite{R6503,R6535} suggest that 
this is the right order of
magnitude for the vector and scalar interaction coefficients. Here, we
emphasize that the existing experimental results on the 
$B \rar K^\ast \ell^+ \ell^-$ and $B \rar K \ell^+ \ell^-$ decays put
stronger restrictions on some of the new Wilson coefficients. For example,
$-2 \le C_{LL} \le 0$, $0 \le C_{RL} \le 2.3$, $-1.5 \le C_{T} \le 1.5$ and
$-3.3 \le C_{TE} \le 2.6$, and all of the remaining Wilson coefficients vary
in the region $- \vel C_{10}^{SM} \ver \le C_X \le \vel C_{10}^{SM} \ver$.

In Figs. 1(2), the dependence of the ${\cal A}_{FB}^{LL}$ on $q^2$
for the $\Lambda_b \rar \Lambda \mu^+ \mu^-$ decay at five different
values of $C_{LL}(C_{LR})$, are presented. We observe from these figures
that zero position of ${\cal A}_{FB}^{LL}$ is shifted compared to that of
the SM result, and this behavior of ${\cal A}_{FB}^{LL}$ is very similar
for both new Wilson coefficients $C_{LL}$ and $C_{LR}$. When both these
coefficients get positive (negative) values, the zero position of 
${\cal A}_{FB}^{LL}$ shifts to the left(right) compared to that of the SM
case.

It should be noted here that, for the above--mentioned cases 
${\cal A}_{FB}^{LL}$ passes through zero for $q^2 < 7~GeV^2$. Therefore,
this zero position of  ${\cal A}_{FB}^{LL}$ is free from long distance
$J/\psi$ contributions.

Our detailed numerical analysis shows that, the zero position of
${\cal A}_{FB}^{LL}$ for the $\Lambda_b \rar \Lambda \mu^+ \mu^-$ decay
is practically independent of the tensor type interactions, and also we
observe that the value of the forward--backward asymmetry is quite small for
the scalar type interactions. Therefore, we do not present the dependence of
${\cal A}_{FB}^{LL}$ on $q^2$ at fixed values of scalar and tensor
interaction coefficients. Moreover, for the 
$\Lambda_b \rar \Lambda \mu^+ \mu^-$ decay, the dependencies of 
${\cal A}_{FB}^{LT}$ and ${\cal A}_{FB}^{TL}$ on $q^2$ are very similar to 
that of  ${\cal A}_{FB}^{LL}$ case, i.e., zero position of  
${\cal A}_{FB}^{LT}$ is shifted to the right(left) compared to that of the 
SM result when the new Wilson coefficients $C_{LL}$ and $C_{LR}$ are 
negative(positive).

The zero position of the forward--backward asymmetry for the
$\Lambda_b \rar \Lambda \tau^+ \tau^-$ decay is absent for all
Wilson coefficients (except the $q^2_{max} = (m_{\Lambda_b} - m_\Lambda)^2$
point), and hence it is insensitive to the
new physics beyond the SM. But, it should be noted here that the values of 
${\cal A}_{FB}^{LT}$ and ${\cal A}_{FB}^{TL}$ at all $q^2$ have opposite 
sign in the presence of tensor interaction, when compared to that of the SM
result. Therefore, if ${\cal A}_{FB}^{LT}$ and ${\cal A}_{FB}^{TL}$ 
are measured far from zero position in the future--planned experiments,
the sign of these asymmetries can give unambiguous information about new
physics beyond the SM, more precisely, about the existence of the tensor
type interaction (see Figs. (3) and (4)).

We can get additional information by studying the dependence of
${\cal A}_{FB}^{NN}$ (and ${\cal A}_{FB}^{TT}= - {\cal A}_{FB}^{NN}$) on
$q^2$. First of all, this asymmetry gets positive(negative) value when
$C_{T}$ is negative(positive) compared to the same case in the SM. The
behavior of ${\cal A}_{FB}^{NN}$ is opposite to the above one in the
presence of $C_{TE}$. Remember that ${\cal A}_{FB}^{NN}\approx 0$ in
the SM (see Fig. (5)). Also, ${\cal A}_{FB}^{NN}$ is very sensitive to the
presence of the scalar interactions. When Wilson coefficients of scalar
interactions, $C_{LRRL}$ and $C_{LRLR}$, are negative(positive), the value of
${\cal A}_{FB}^{NN}$ becomes positive(negative) for the
$\Lambda_b \rar \Lambda \tau^+ \tau^-$ decay (see Fig.(6)).

Obviously, it follows from the explicit expressions of ${\cal A}_{FB}^{ij}$  
that they all depend both on $q^2$ and the new Wilson coefficients.
Therefore there may appear difficulties in simultaneous study of the 
dependence of the physical observables on both parameters. In order to avoid
such difficulties, we must eliminate the dependence of ${\cal A}_{FB}^{ij}$
on one of these parameters. In the present work, the $q^2$
dependence of ${\cal A}_{FB}^{ij}$ is eliminated by performing integration
over $q^2$ in the allowed kinematical region, i.e., we average the polarized
${\cal A}_{FB}^{ij}$ which is defined as
\bea
\label{e6523}
\lla {\cal A}_{FB}^{ij} \rra = \frac{\ds 
\int_{4 m_\ell^2}^{(m_{\Lambda_b} - m_\Lambda)^2} 
{\cal A}_{FB}^{ij} \frac{d {\cal B}}{dq^2} dq^2}
{\ds \int_{4 m_\ell^2}^{(m_{\Lambda_b} - m_\Lambda)^2}      
\frac{d {\cal B}}{dq^2} dq^2}~.
\eea  

In Fig.(7) we depict the dependence of $\lla {\cal A}_{FB}^{LL} \rra$ on
$C_X$ for the $\Lambda_b \rar \Lambda \mu^+ \mu^-$ decay. The intersection
of all curves corresponds to the SM case. It follows from this figure that,
$\lla {\cal A}_{FB}^{LL} \rra$ has symmetric behavior in its dependence on
the tensor type interactions and except $C_{RR}$ it remains smaller compared
to the SM result at negative values of all type of interactions. At
positive values of the new Wilson coefficients  
$\lla {\cal A}_{FB}^{LL} \rra > \lla {\cal A}_{FB}^{LL} \rra_{SM}$ for all
type of scalar interactions and for the vector interactions 
$C_{LL}$, $C_{LR}$ and $C_{RR}$.

Depicted in Fig. (8) is the dependence of $\lla {\cal A}_{FB}^{LT} \rra$ on
$C_X$ for the $\Lambda_b \rar \Lambda \mu^+ \mu^-$ decay. We observe from
this figure that $\lla {\cal A}_{FB}^{LT} \rra$ depends more strongly on the
tensor interaction coefficients $C_{T}$ and $C_{TE}$. When $C_{T}<0$,
$\lla {\cal A}_{FB}^{LT} \rra$ is positive and when $-4 < C_{T}< -2$ , 
$\lla {\cal A}_{FB}^{LT} \rra$ reaches its maximum value (about $\sim 5\%$).
On the other hand when $C_{T}$ is positive, $\lla {\cal A}_{FB}^{LT} \rra$
changes sign and gets negative values. The situation is different for the
other tensor interaction coefficient, namely, $C_{TE}$. When 
$-4 < C_{TE} <-2$, $\lla {\cal A}_{FB}^{LT} \rra$ is positive, and when
$-2 < C_{TE} <0$, $\lla {\cal A}_{FB}^{LT} \rra$ is negative.
$\lla {\cal A}_{FB}^{LT} \rra$ attains at positive values again when 
$C_{TE}>0$. Therefore measurement of $\lla {\cal A}_{FB}^{LT} \rra$ in
experiments can give essential information about the existence of tensor
interactions.

The dependence of $\lla {\cal A}_{FB}^{TL} \rra$ on the new Wilson
coefficients, for the $\Lambda_b \rar \Lambda \mu^+ \mu^-$ decay, 
is vice versa of the behavior of $\lla {\cal A}_{FB}^{LT}
\rra$ explained above (see Fig. (9)).

In Fig. (10) we present the dependence of $\lla {\cal A}_{FB}^{LL} \rra$
on the new Wilson coefficients for the $\Lambda_b \rar \Lambda \tau^+ \tau^-$
decay. From this figure one can easily conclude that $\lla {\cal
A}_{FB}^{LL} \rra$ is sensitive to the presence of the new Wilson
coefficients, except $C_{RR}$, $C_{LL}$, $C_{LRLR}$, $C_{LRRL}$ and
$C_{RLLR}$. Practically, at all negative(positive) values of the new Wilson
coefficients (except the regions $-1 \le C_{T} \le 0$ and 
$0 \le C_{TE} \le 0.4$) $\lla {\cal A}_{FB}^{LL} \rra$ is smaller(larger)
compared to $\lla {\cal A}_{FB}^{LL} \rra_{SM}$ predicted by the SM.
Along the same lines, the dependence of $\lla {\cal A}_{FB}^{LT} \rra$ for the
$\Lambda_b \rar \Lambda \tau^+ \tau^-$ decay is presented in Fig. (11). In
this case $\lla {\cal A}_{FB}^{LT} \rra$ seems to be strongly dependent on 
$C_{T}$, $C_{TE}$, $C_{LR}$ and $C_{LRRL}$. In the case of tensor
interaction, at all values of $C_{T}$($C_{TE}$) we observe that
$\lla {\cal A}_{FB}^{LT} \rra > \lla {\cal A}_{FB}^{LT} \rra_{SM}$, while
for the coefficient $C_{LR}$, the magnitude of $\lla {\cal A}_{FB}^{LT}
\rra$ is smaller(larger) compared to the SM result when $C_{LR}$ gets
negative(positive) values. For the scalar type interaction induced by 
$C_{LRRL}$, the situation is vice versa when compared to the $C_{LR}$ case.  

As far as the dependence of $\lla {\cal A}_{FB}^{TL} \rra$ on $C_X$
is concerned, apart from an overall sign, practically, its behavior is almost 
the same as that of $\lla {\cal A}_{FB}^{LT} \rra$ for the $\Lambda_b \rar
\Lambda \tau^+ \tau^-$ decay (see Figs. (11) and (12)). 
  
In Fig. (13) we present the dependence of $\lla {\cal A}_{FB}^{NN} \rra = 
- \lla {\cal A}_{FB}^{TT} \rra$ on the Wilson coefficients for the
$\Lambda_b \rar \Lambda \tau^+ \tau^-$ decay, and we observe that it is
strongly dependent on $C_{T}$, $C_{TE}$ and the scalar type interactions
with the coefficients $C_{LRLR}$ and $C_{LRRL}$.

From these results it follows that several of the double--lepton
polarization forward--backward 
asymmetries demonstrate sizable departure from the SM results and they are
sensitive to the existence of different types of new interactions. For this
reason, 
study of these observables can be very useful in looking for new physics
beyond the SM. 

At the end of this section we want to discuss the following problem.
Obviously, if new physics beyond the SM exists, it effects not only the
polarized ${\cal A}_{FB}$, but also the branching ratio. The measurement of
the branching ratio in experiments is easier and therefore its investigation
is more convenient for establishing new physics. The main question is,
whether there could appear situations in which the value 
of the branching ratio coincides with that of the SM result, while polarized
${\cal A}_{FB}$ does not. To find out an answer to this question we study the   
correlation between the averaged, polarized $\lla {\cal A}_{FB} \rra$ and
the branching ratio. In further analysis we vary the branching ratio of the   
$\Lambda_b \rar \Lambda \mu^+ \mu^-~(\Lambda \tau^+ \tau^-)$ decay 
between the values 
$(3-6)\times 10^{-6}~[(3-6)\times 10^{-7}]$, which is very close to the SM
results.

Our comment on the $\Lambda_b \rar \Lambda \mu^+ \mu^-$ decay, as far as 
the above--mentioned correlated relation is concerned, can briefly be
summarized as follows 
(remember that, the intersection of all curves corresponds to the SM value):

\begin{itemize}
\item for $\lla {\cal A}_{FB}^{LL} \rra$, $\lla {\cal A}_{FB}^{LT} \rra$ and
$\lla {\cal A}_{FB}^{TL} \rra$, such a region exists only for 
$C_{LR}$.
\end{itemize}

The situation is much richer for the $\Lambda_b \rar \Lambda \tau^+ \tau^-$
decay. In Figs. (14)--(17), we depict the dependence of the averaged,
forward--backward polarized asymmetries $\lla {\cal A}_{FB}^{LL} \rra$; 
$\lla {\cal A}_{FB}^{LT} \rra \approx - \lla {\cal A}_{FB}^{TL} \rra$; 
$\lla {\cal A}_{FB}^{NT} \rra = \lla {\cal A}_{FB}^{TN} \rra$; 
$\lla {\cal A}_{FB}^{NN} \rra = - \lla {\cal A}_{FB}^{TT} \rra$, on 
branching ratio. It follows from these figures that, indeed, there exist
certain regions of the new Wilson coefficients in where the study of the
doubly polarized ${\cal A}_{FB}$ can establish new
physics beyond the SM.

In conclusion, we present the analysis for the forward--backward asymmetries
using a general, model independent form of the effective Hamiltonian. We obtain 
that the determination of the zero position of $\lla {\cal A}_{FB}^{LL} \rra$ 
can serve as a good probe for establishing new physics beyond the SM. Finally 
we obtain that there exist certain regions of the new Wilson coefficients for 
which, only study of the polarized forward--backward asymmetry gives invaluable 
information in search of new physics beyond the SM.

\newpage

\section*{Figure captions}
{\bf Fig. (1)} The dependence of the double--lepton polarization asymmetry
${\cal A}_{FB}^{LL}$ on $q^2$ at four fixed
values of $C_{LL}$, for the $\Lambda_b \rar \Lambda \mu^+ \mu^-$ decay.\\ \\
{\bf Fig. (2)} The same as in Fig. (1), but at four fixed 
values of $C_{LR}$.\\ \\
{\bf Fig. (3)} The dependence of the double--lepton polarization asymmetry
${\cal A}_{FB}^{LT}$ on $q^2$ at four fixed 
values of $C_{T}$, for the $\Lambda_b \rar \Lambda \tau^+ \tau^-$ decay.\\ \\
{\bf Fig. (4)} The same as in Fig. (3), but for ${\cal A}_{FB}^{TL}$.\\ \\
{\bf Fig. (5)} The same as in Fig. (3), but for ${\cal A}_{FB}^{NN}$.\\ \\
{\bf Fig. (6)} The same as in Fig. (5), but at four fixed 
values of $C_{LRRL}$.\\ \\
{\bf Fig. (7)} The dependence of the averaged forward--backward
double--lepton polarization asymmetry $\lla {\cal A}_{FB}^{LL} \rra$
on the new Wilson coefficients $C_X$, for the $\Lambda_b \rar \Lambda \mu^+ \mu^-$
decay.\\ \\
{\bf Fig. (8)} The same as in Fig. (7), but for ${\cal A}_{FB}^{LT}$.\\ \\
{\bf Fig. (9)} The same as in Fig. (8), but for the averaged 
forward--backward double--lepton polarization asymmetry
$\lla {\cal A}_{FB}^{TL} \rra$.\\ \\
{\bf Fig. (10)} The same as in Fig. (7), but for the 
$\Lambda_b \rar \Lambda \tau^+ \tau^-$ decay.\\ \\
{\bf Fig. (11)} The same as in Fig. (8), but for the 
$\Lambda_b \rar \Lambda \tau^+ \tau^-$ decay.\\ \\
{\bf Fig. (12)} The same as in Fig. (9), but for the 
$\Lambda_b \rar \Lambda \tau^+ \tau^-$ decay.\\ \\
{\bf Fig. (13)} The same as in Fig. (12), but for the 
$\lla {\cal A}_{FB}^{NN} \rra = - \lla {\cal A}_{FB}^{TT} \rra$.\\ \\
{\bf Fig. (14)}  Parametric plot of the correlation between the averaged
forward--backward double--lepton polarization asymmetry
$\lla {\cal A}_{FB}^{LL} \rra$ and the branching ratio for the
$\Lambda_b \rar \Lambda \tau^+ \tau^-$ decay.\\ \\
{\bf Fig. (15)} The same as in Fig. (14), but for the 
the correlation between the averaged
forward--backward double--lepton polarization asymmetry
$\lla {\cal A}_{FB}^{LT} \rra$ and the branching ratio.\\ \\
{\bf Fig. (16)} The same as in Fig. (15), but for the 
the correlation between the averaged
forward--backward double--lepton polarization asymmetry
$\lla {\cal A}_{FB}^{NT} \rra = \lla {\cal A}_{FB}^{TN} \rra$ and 
the branching ratio.\\ \\
{\bf Fig. (17)} The same as in Fig. (16), but for the 
the correlation between the averaged
forward--backward double--lepton polarization asymmetry
$\lla {\cal A}_{FB}^{NN} \rra = - \lla {\cal A}_{FB}^{TT} \rra$ and 
the branching ratio.\\ \\

\newpage

\begin{figure}
\vskip 1.5 cm
    \includegraphics{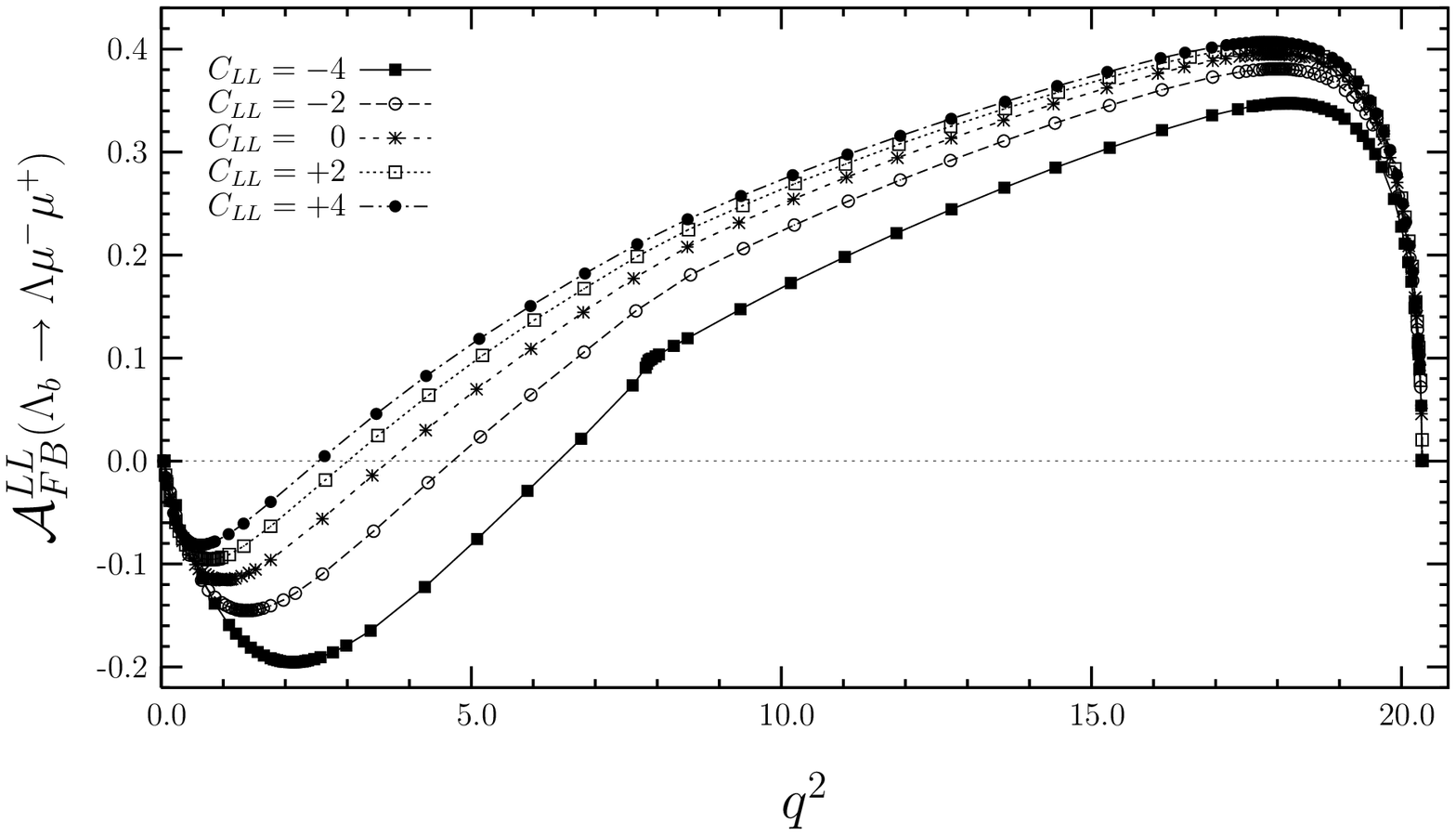}
\vskip 7.8cm
\caption{}
\end{figure}

\begin{figure}
\vskip 2.5 cm
    \includegraphics{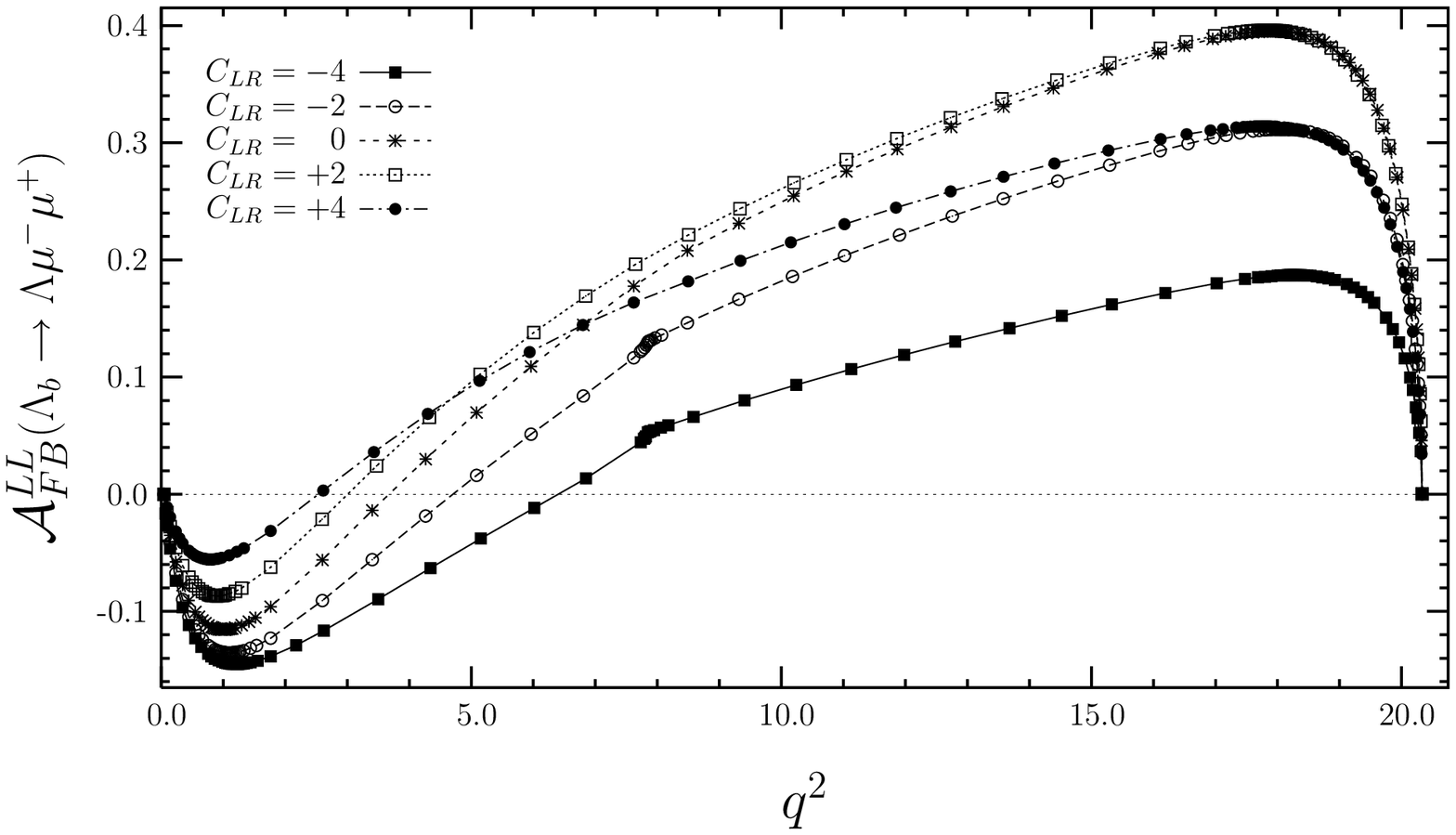}
\vskip 7.8 cm
\caption{}
\end{figure}

\begin{figure}
\vskip 1.5 cm
    \includegraphics{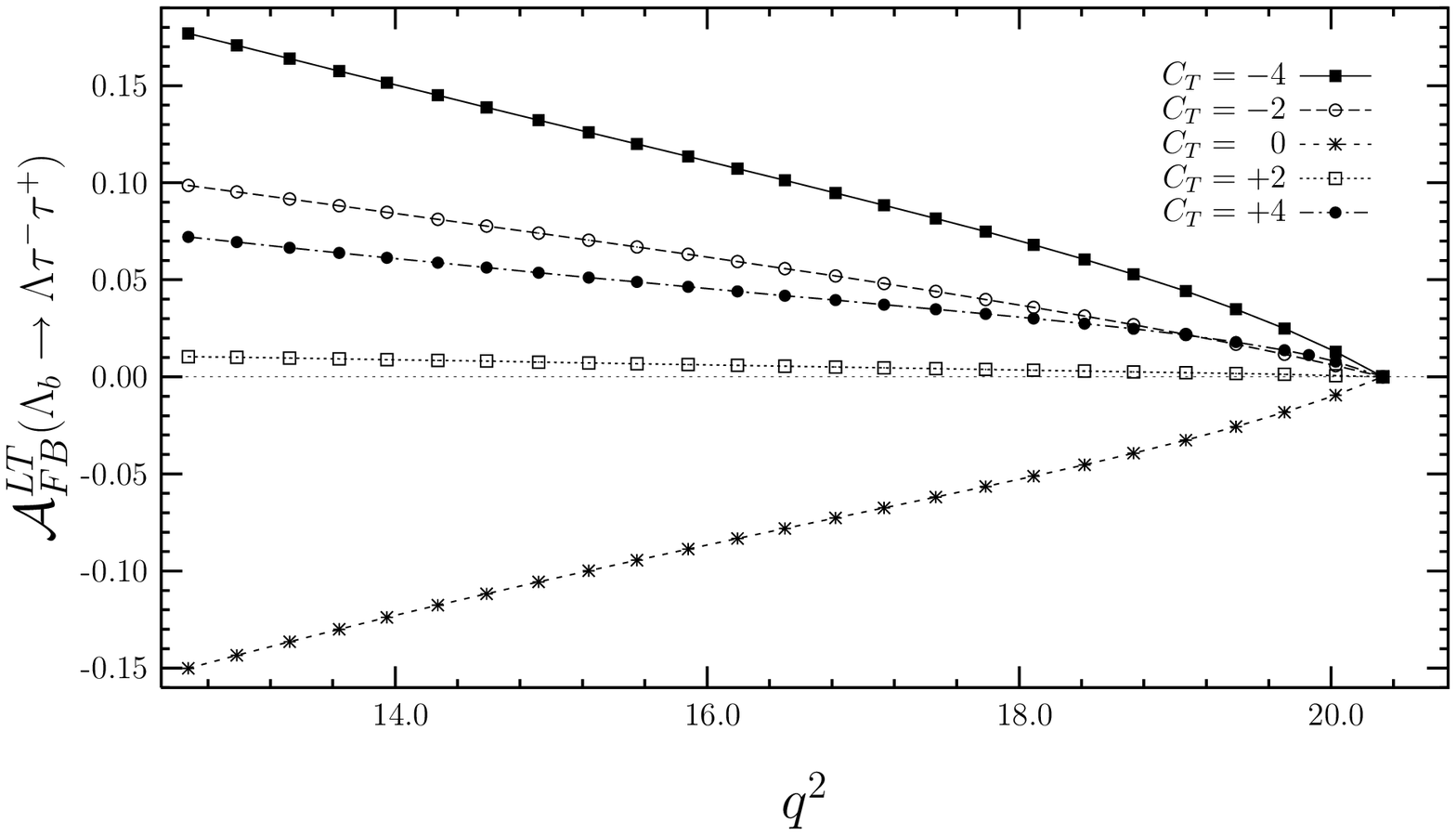}
\vskip 7.8cm
\caption{}
\end{figure}

\begin{figure}
\vskip 2.5 cm
    \includegraphics{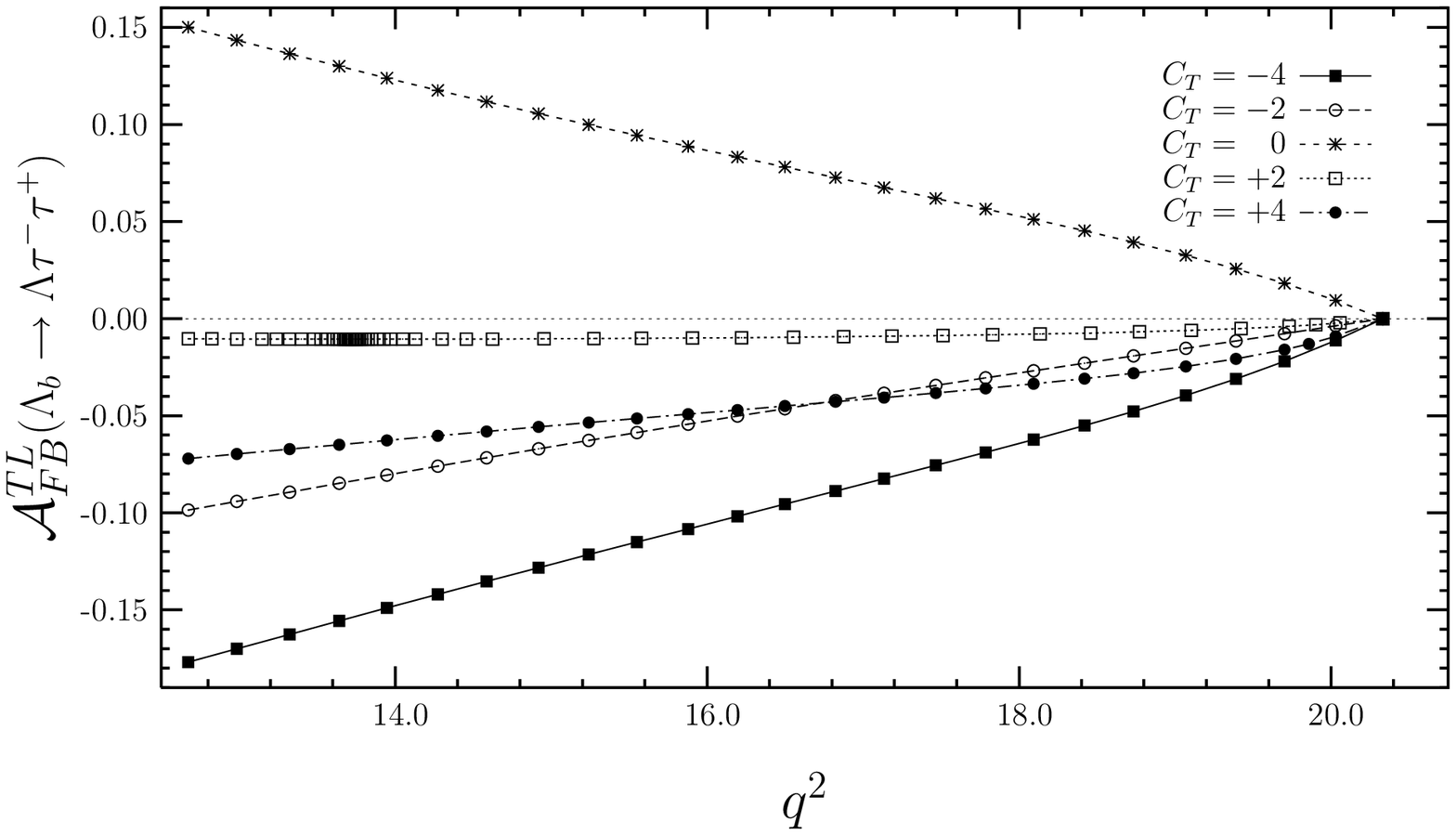}
\vskip 7.8 cm
\caption{}
\end{figure}

\begin{figure}
\vskip 2.5 cm
    \includegraphics{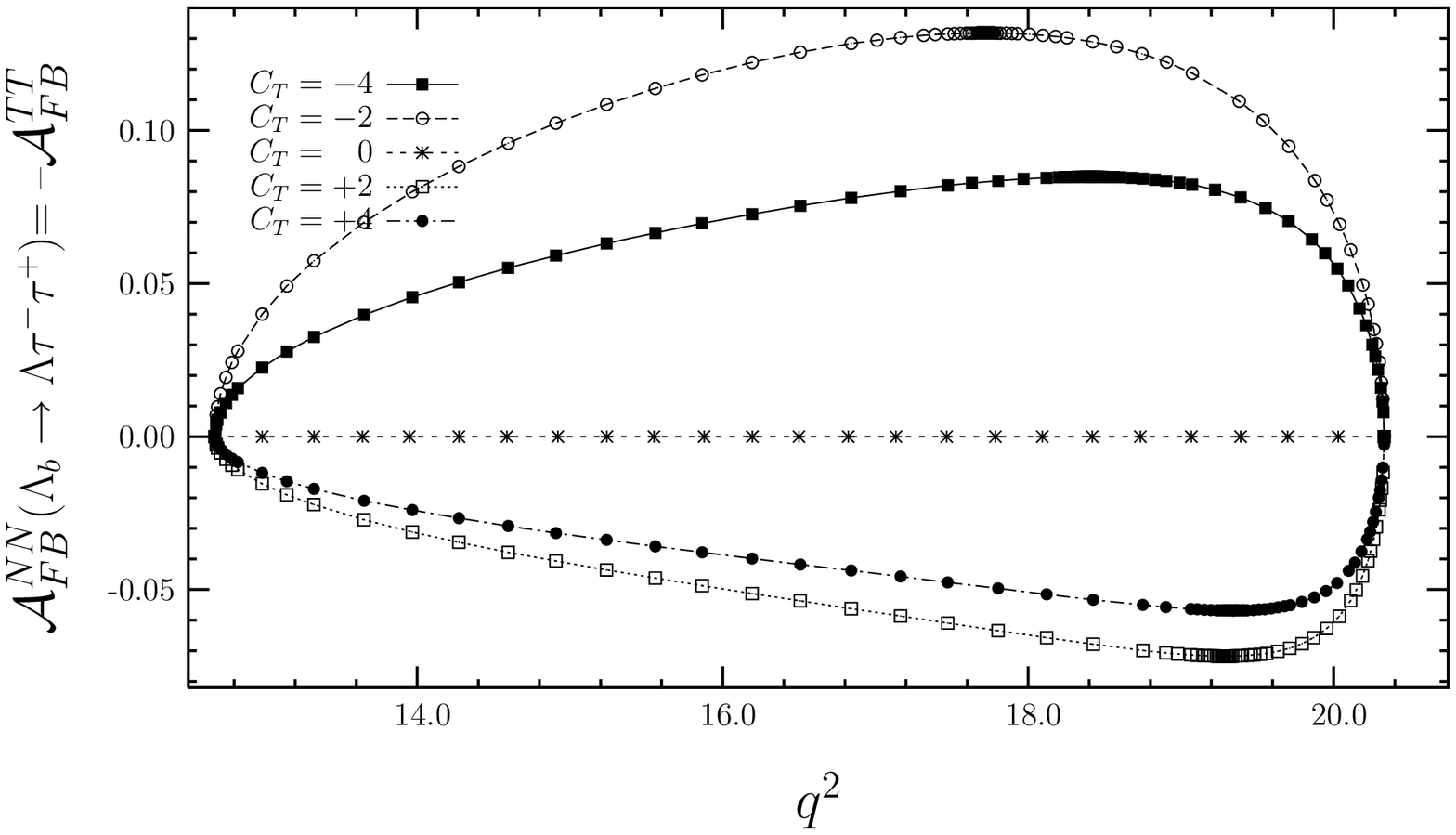}
\vskip 7.8 cm
\caption{}
\end{figure}

\begin{figure}
\vskip 1.5 cm
    \includegraphics{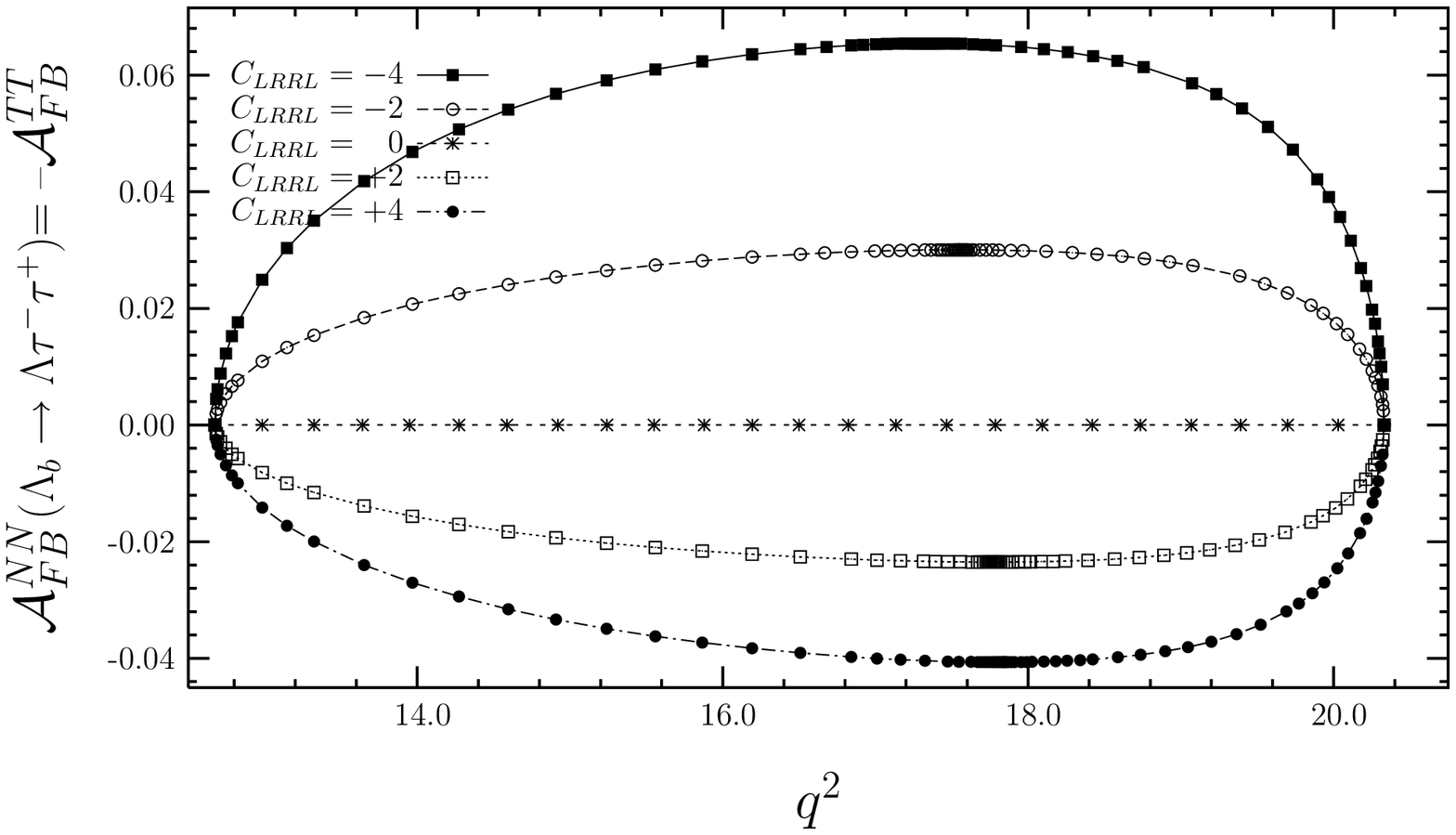}
\vskip 7.8cm
\caption{}
\end{figure}

\begin{figure}
\vskip 2.5 cm
    \includegraphics{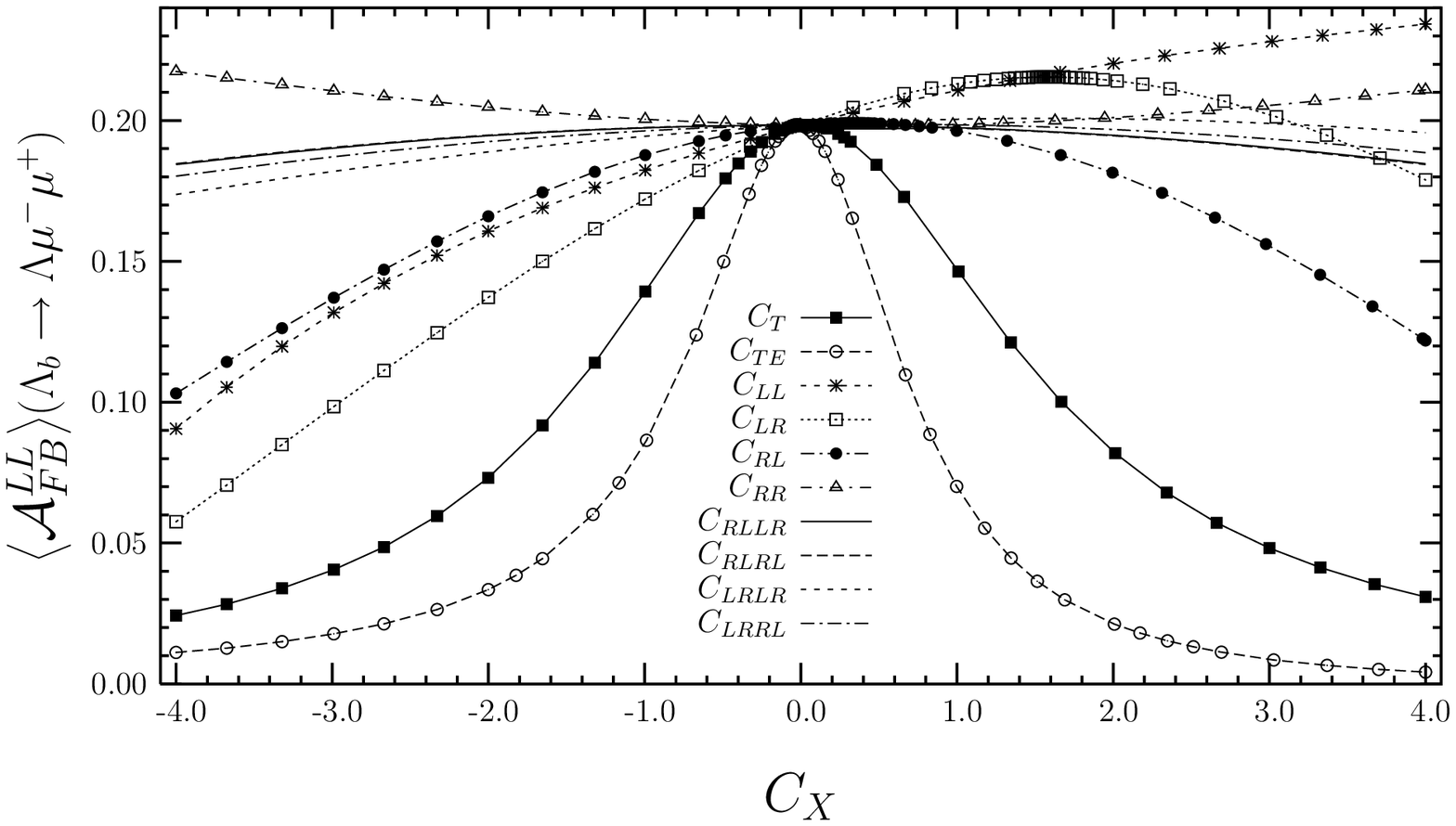}
\vskip 7.8 cm
\caption{}
\end{figure}

\begin{figure}
\vskip 1.5 cm
    \includegraphics{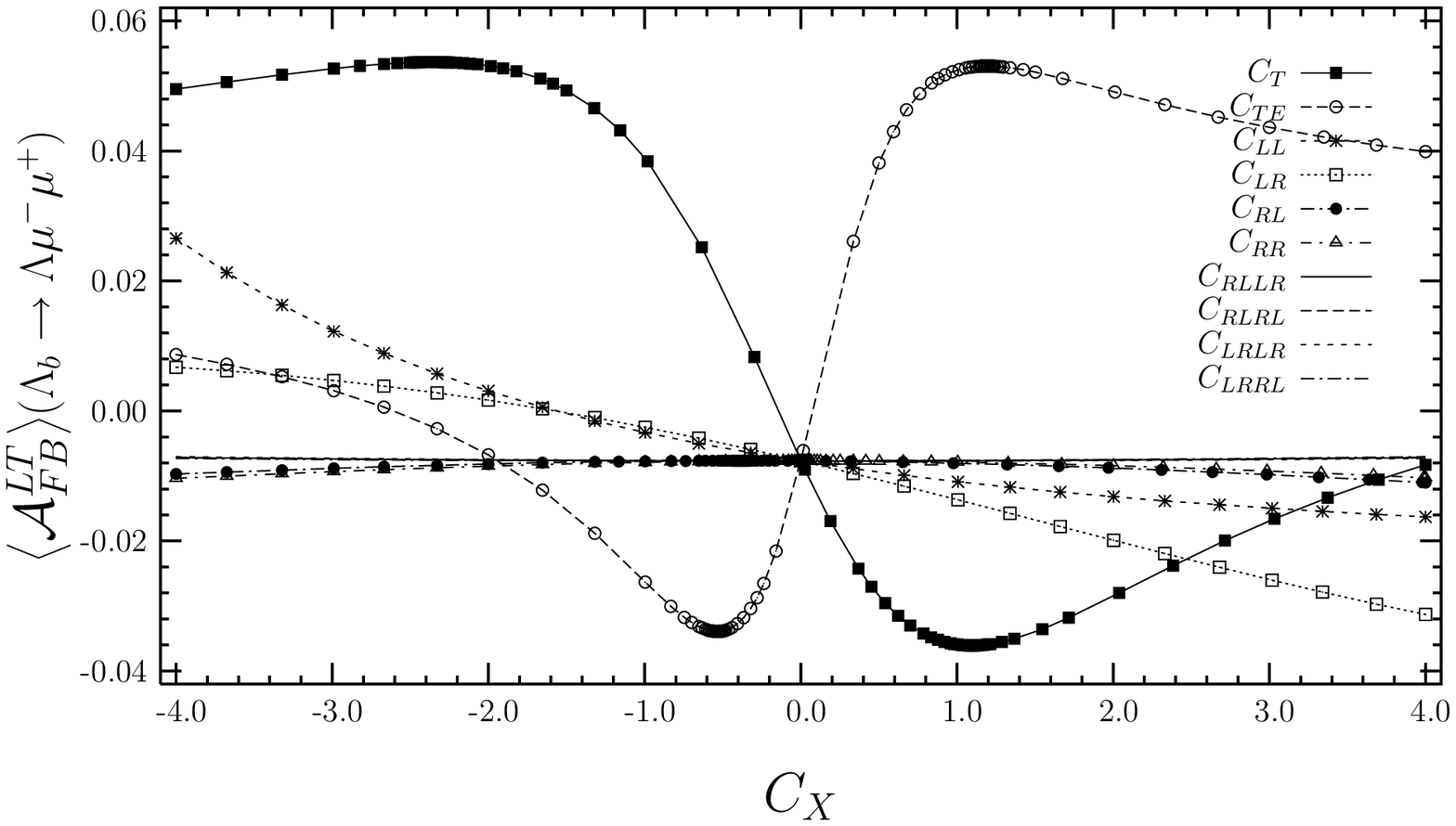}
\vskip 7.8cm
\caption{}
\end{figure}

\begin{figure}
\vskip 2.5 cm
    \includegraphics{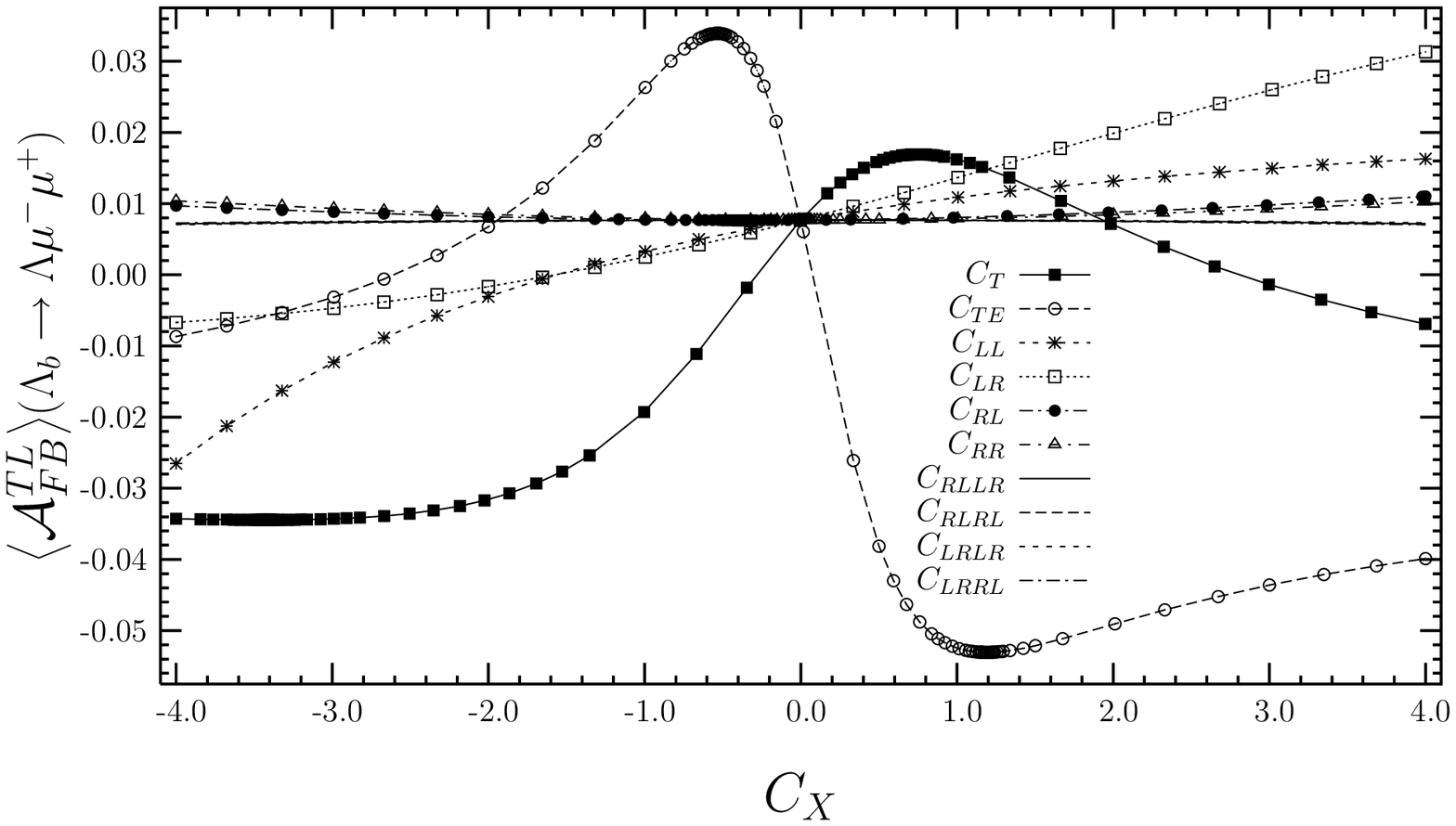}
\vskip 7.8 cm
\caption{}
\end{figure}

\begin{figure}
\vskip 1.5 cm
    \includegraphics{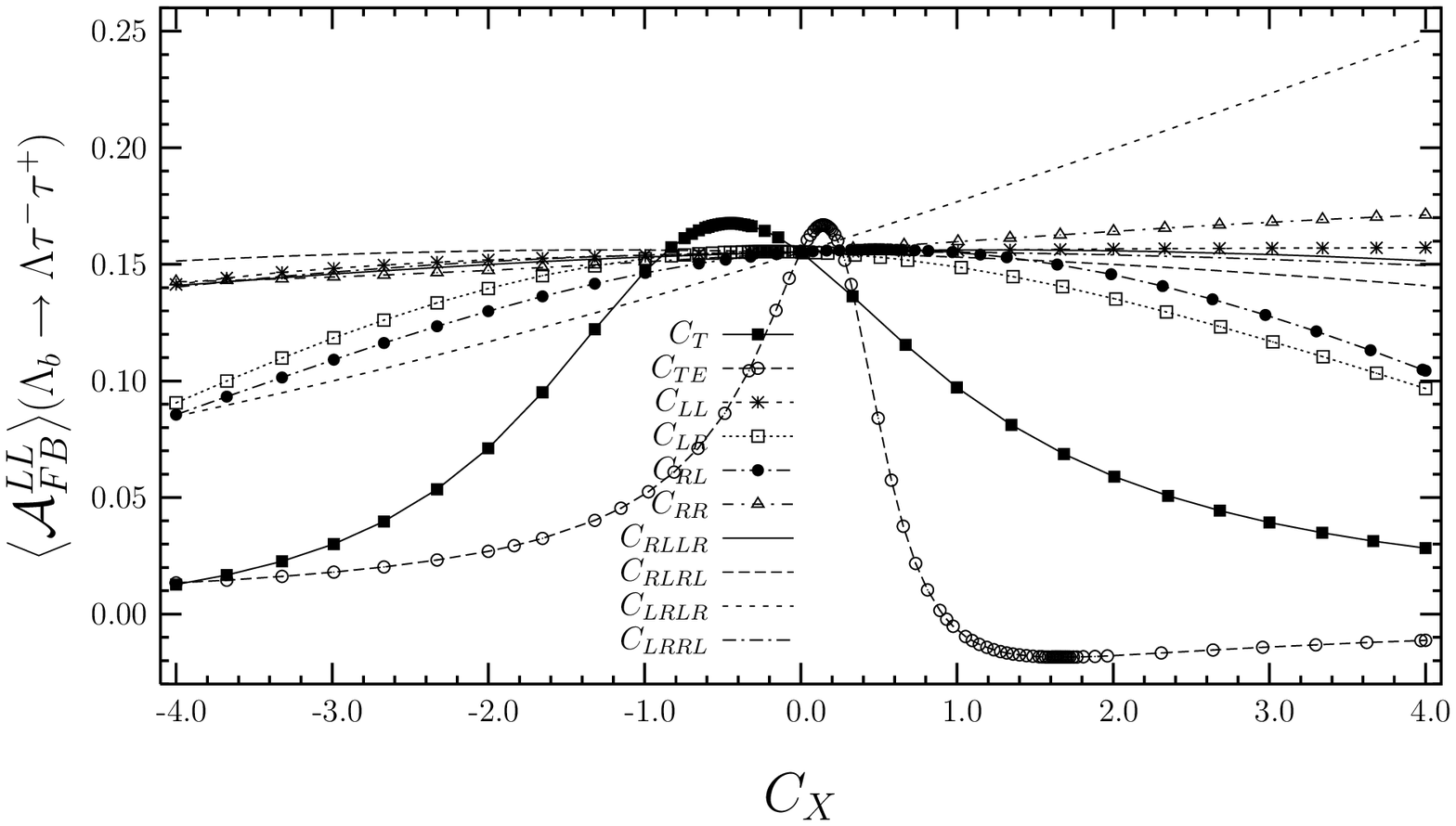}
\vskip 7.8cm
\caption{}
\end{figure}

\begin{figure}
\vskip 2.5 cm
    \includegraphics{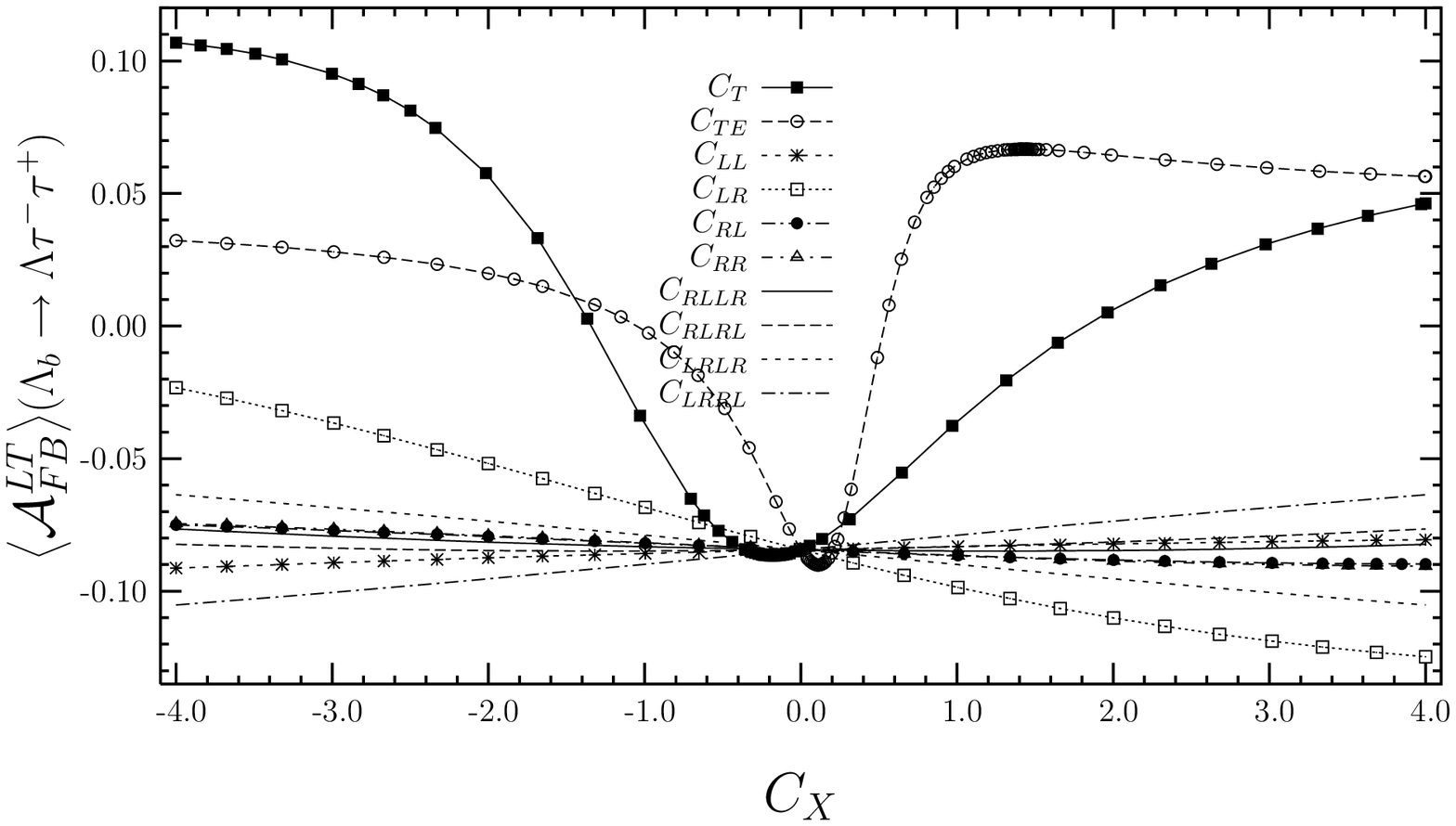}
\vskip 7.8 cm
\caption{}
\end{figure}

\begin{figure}
\vskip 1.5 cm
    \includegraphics{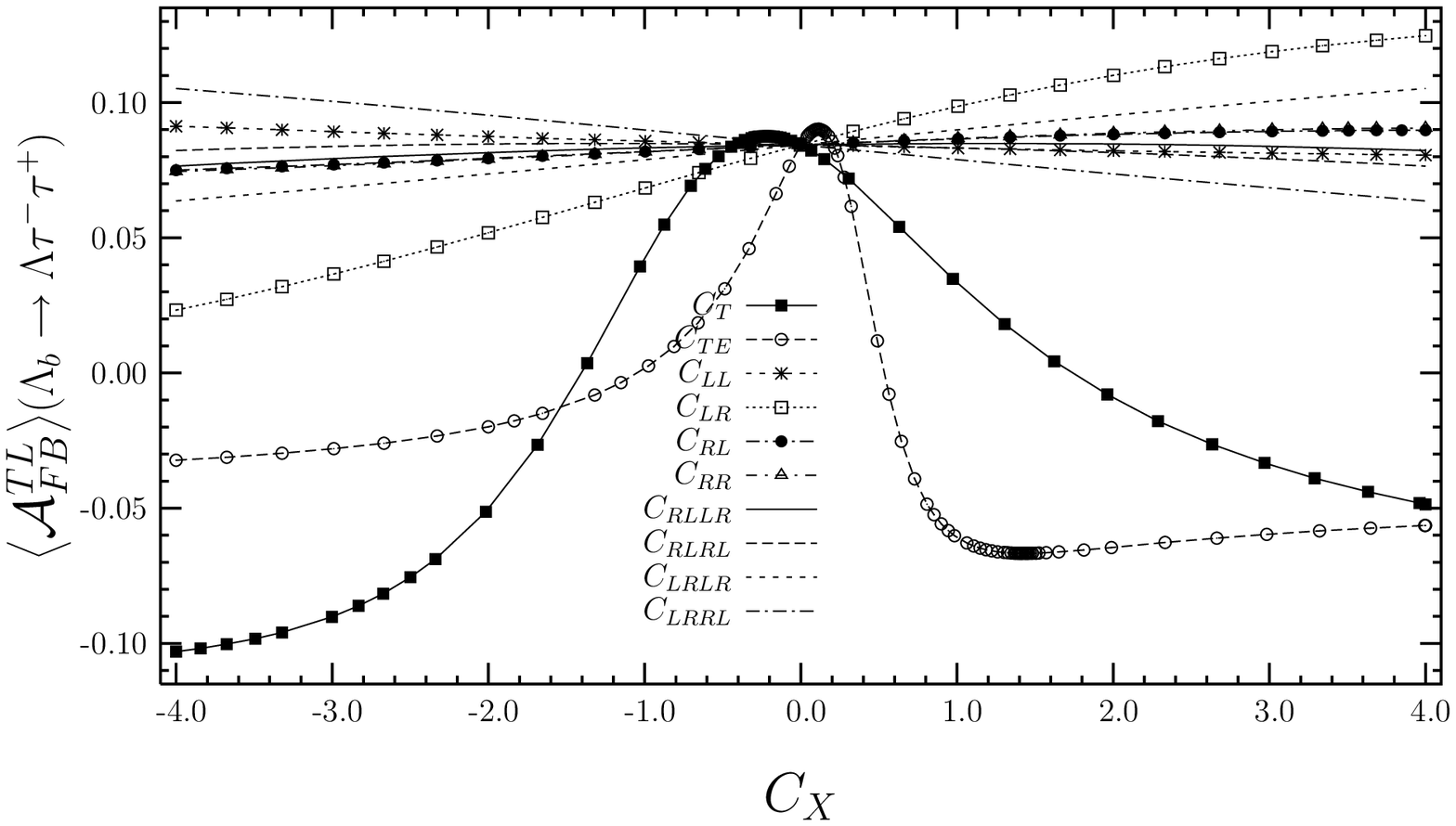}
\vskip 7.8cm
\caption{}
\end{figure}

\begin{figure}
\vskip 1.5 cm
    \includegraphics{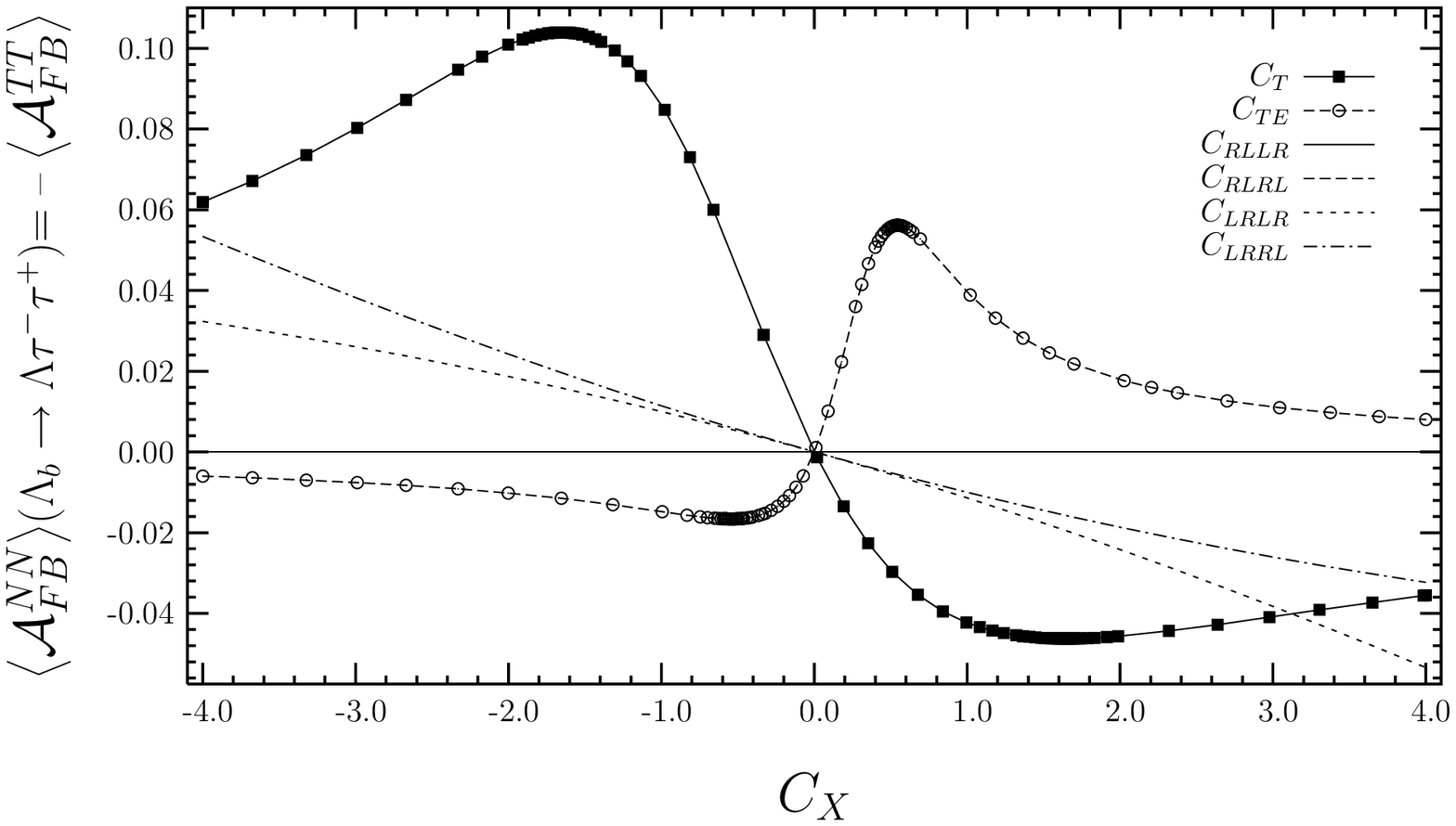}
\vskip 8.8 cm
\caption{}
\end{figure}

\begin{figure}
\vskip 2.5 cm
    \includegraphics{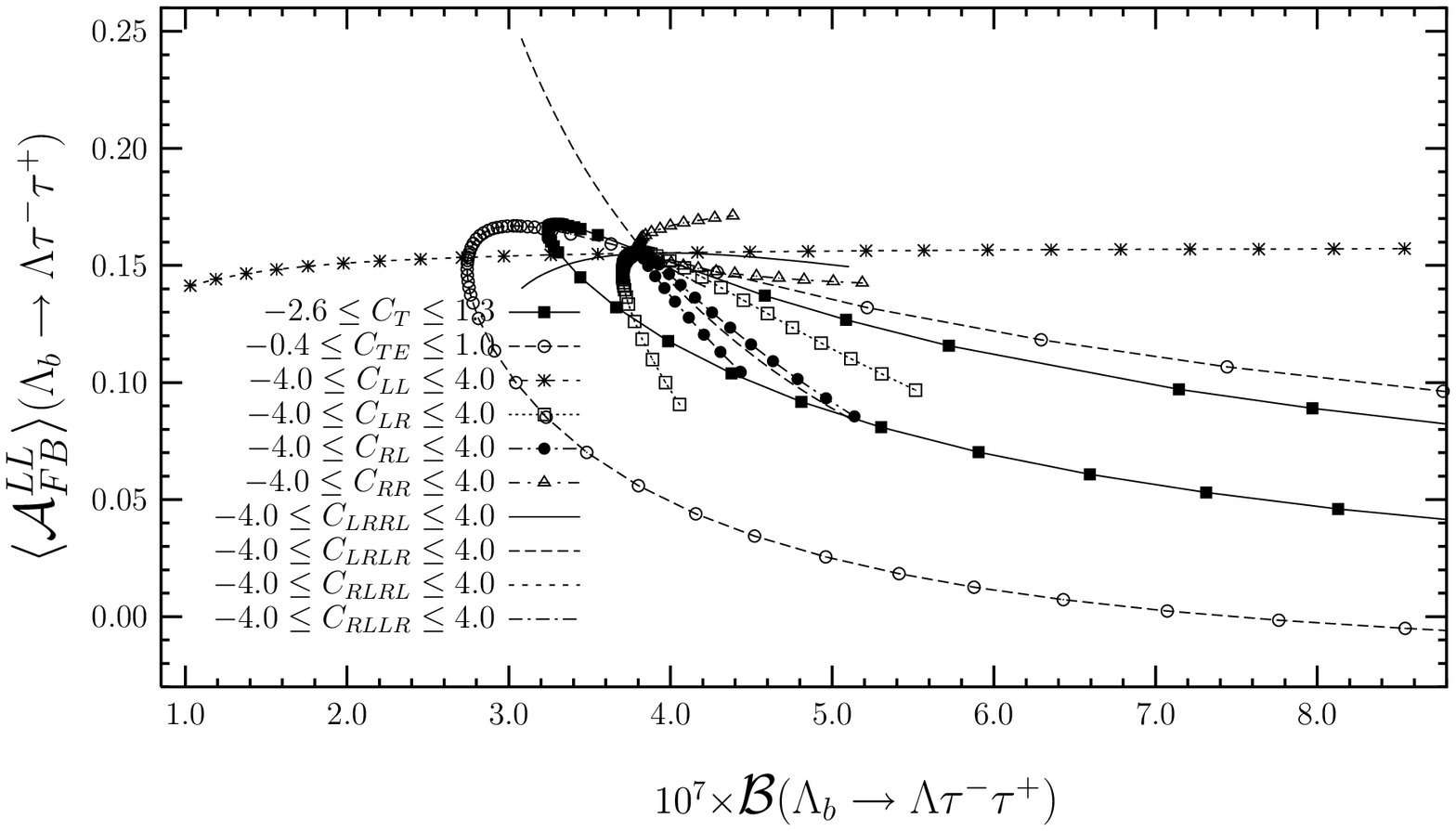}
\vskip 7.8cm
\caption{}
\end{figure}

\begin{figure}
\vskip 2.5 cm
    \includegraphics{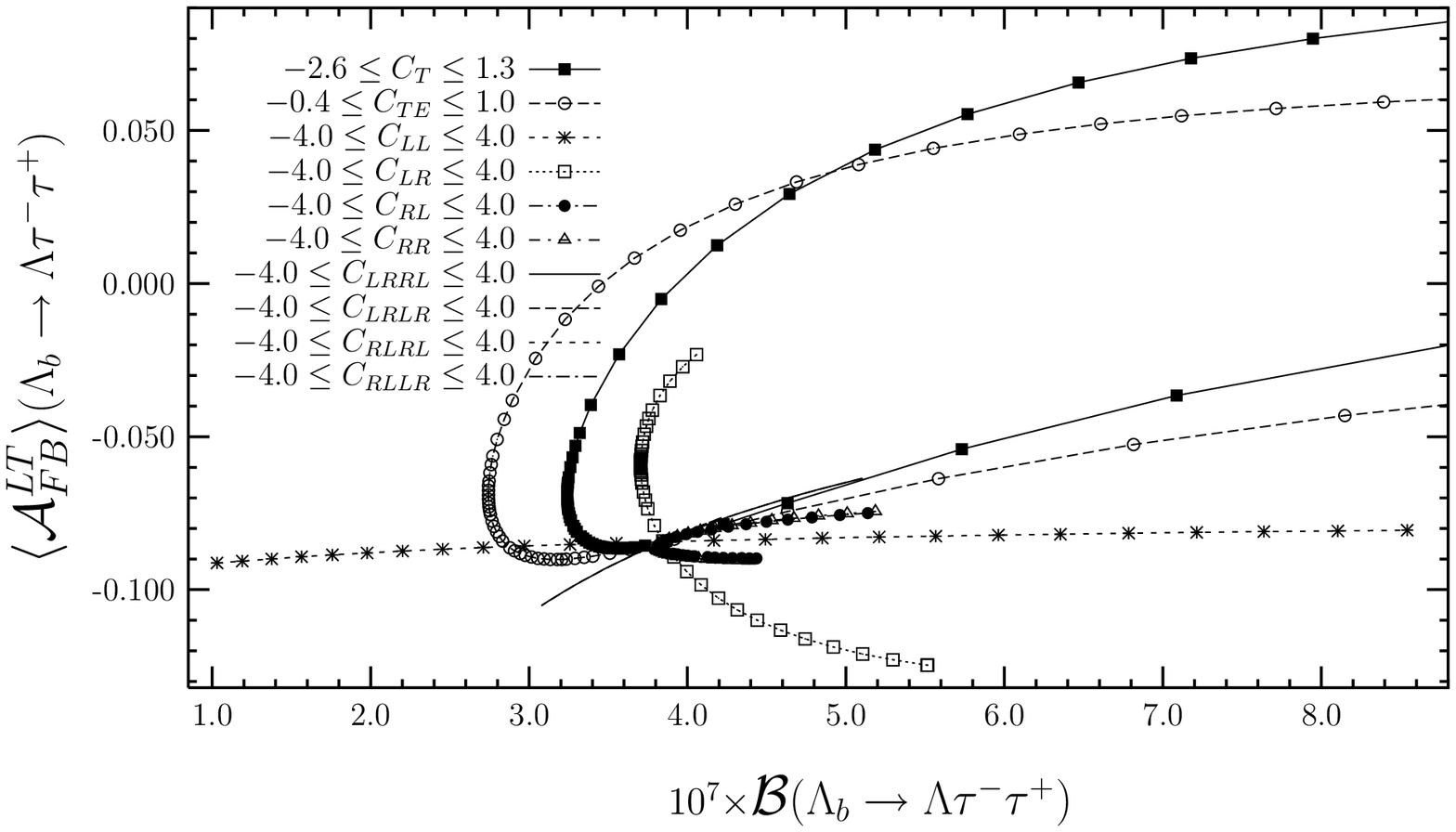}
\vskip 7.8 cm
\caption{}
\end{figure}

\begin{figure}
\vskip 1.5 cm
    \includegraphics{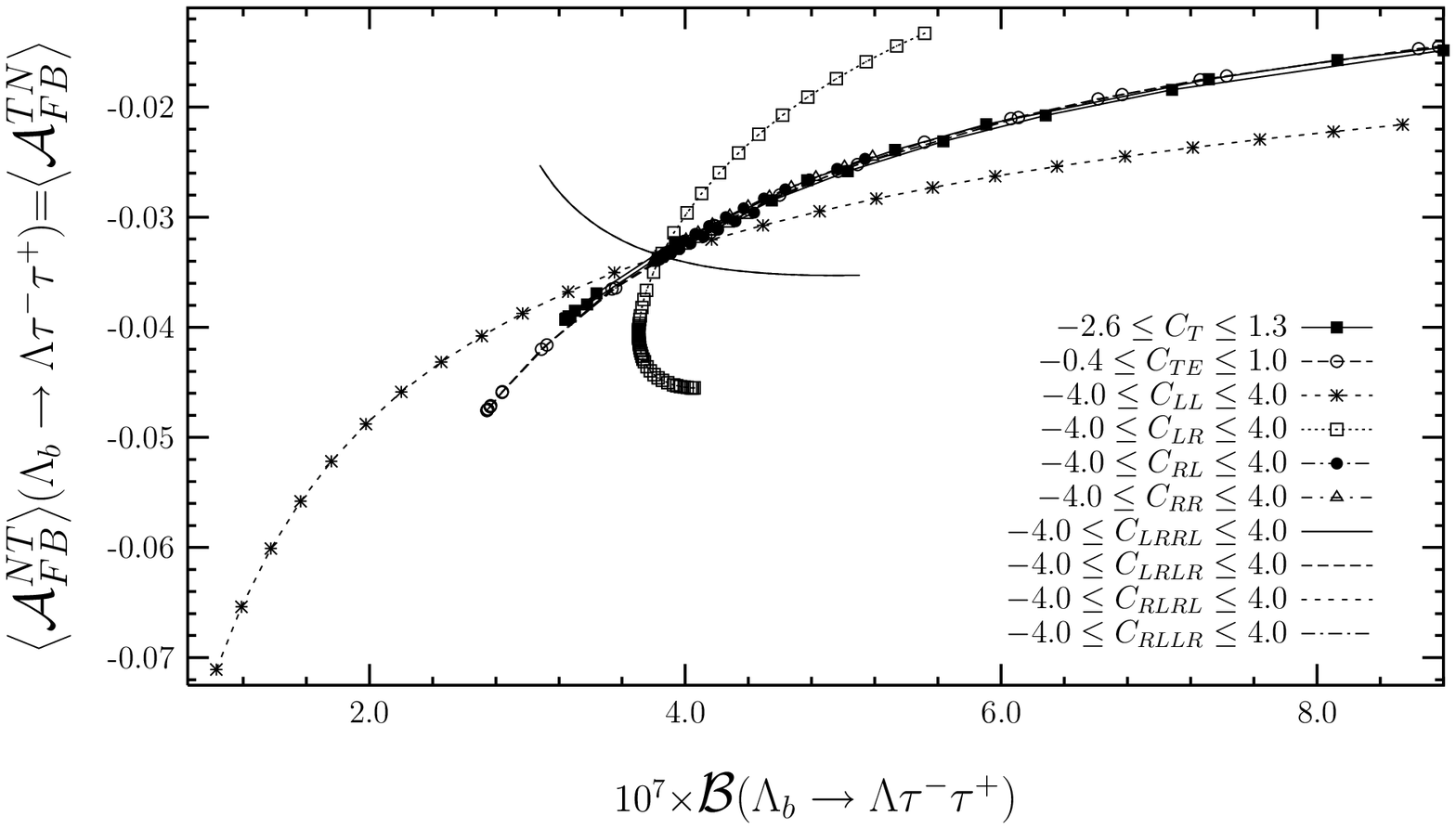}
\vskip 7.8cm
\caption{}
\end{figure}

\begin{figure}
\vskip 2.5 cm
    \includegraphics{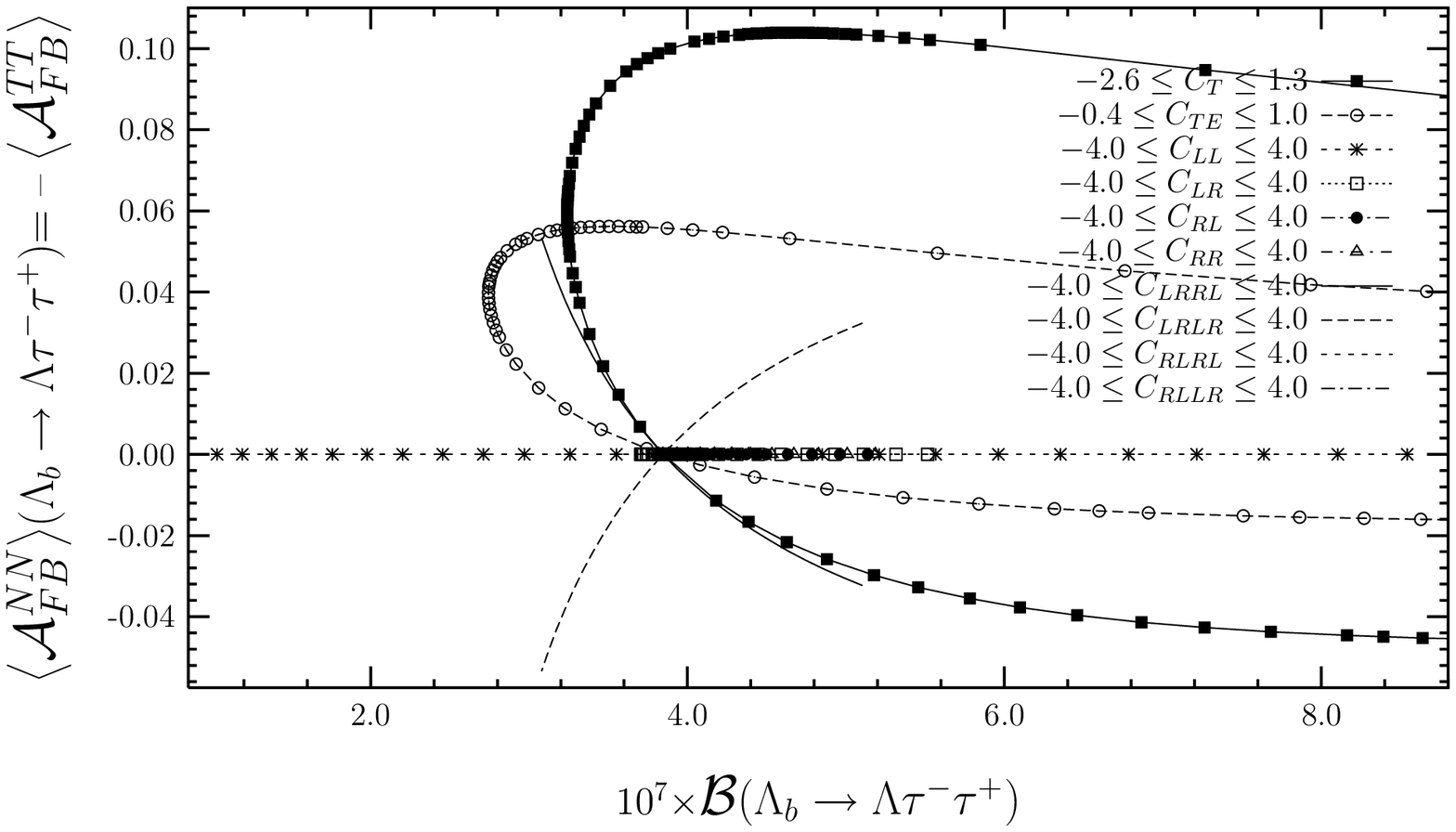}
\vskip 7.8 cm
\caption{}
\end{figure}

\end{document}